\documentclass[a4paper,11pt]{article}
\usepackage{multirow} % Ensure this package is included

\pdfoutput=1 % if your are submitting a pdflatex (i.e. if you have
\usepackage{comment}
\usepackage{jcappub} % for details on the use of the package, please
\usepackage{comment}
\usepackage[T1]{fontenc} % if needed

\newcommand{\bfx}{\mbox{\boldmath$x$}}
\newcommand{\bfk}{\mbox{\boldmath$k$}}

\title{Stage-IV Cosmic Shear with Modified Gravity and Model-independent Screening}

\author[a,1]{M. Tsedrik,\note{Corresponding author.}}
\author[a]{B. Bose,}
\author[a]{P. Carrilho,}
\author[a,b]{A. Pourtsidou,}
\author[c]{S. Pamuk,}
\author[c]{S. Casas,}
\author[c]{J. Lesgourgues}

\affiliation[a]{Institute for Astronomy, University of Edinburgh, \\ Royal Observatory, Blackford Hill, Edinburgh, EH9 3HJ, U.K.}
\affiliation[b]{Higgs Centre for Theoretical Physics, School of Physics and Astronomy, \\ Edinburgh EH9 3FD, UK}
\affiliation[c]{Institute for Theoretical Physics and Cosmology (TTK), \\
RWTH Aachen University, D-52056 Aachen, Germany}

\emailAdd{mtsedrik@ed.ac.uk}

\abstract{We forecast constraints on minimal model-independent parametrisations of several Modified Gravity theories using mock Stage-IV cosmic shear data. We include nonlinear effects and screening, which ensures recovery of General Relativity on small scales. We introduce a power spectrum emulator to accelerate our analysis and evaluate the robustness of the growth index parametrisation with respect to two cosmologies: $\Lambda$CDM and the normal branch of the DGP model. We forecast the uncertainties on the growth index $\gamma$ to be of the order $\sim 10\%$. We find that our halo-model based screening approach demonstrates excellent performance, meeting the precision requirements of Stage-IV surveys. However, neglecting the screening transition results in biased predictions for cosmological parameters. We find that the screening transition shows significant degeneracy with baryonic feedback, requiring a much better understanding of baryonic physics for its detection. Massive neutrinos effects are less prominent and challenging to detect solely with cosmic shear data. 
}

\begin{document}
\maketitle
\flushbottom

\section{Introduction}

Stage-IV cosmological surveys, such as Euclid\footnote{\url{http://euclid-ec.org}}~\citep{Laureijs:2011gra} and the Vera C. Rubin Observatory’s Legacy Survey of Space and Time (LSST)\footnote{\url{https://www.lsst.org/}}~\citep{LSST:2008ijt}, will provide us with high precision data, enabling us to constrain deviations from General Relativity (GR). GR serves as the foundational theory of gravity in the standard cosmological model, $\Lambda$CDM. Enhancing the robustness of our modelling has become imperative, as statistical uncertainties will no longer be the limiting obstacle; instead, Stage-IV data analyses will be limited by the accuracy of the theoretical modelling of the observables.

Cosmic shear, the weak gravitational lensing effect that systematically distorts the shapes of galaxy images, can be directly connected to the spatial distribution of all gravitating matter along the line-of-sight (for comprehensive reviews, see Refs.~\cite{Kilbinger:2014, Bartelmann:2017, Mandelbaum:2018}). For Stage-IV surveys, the small scales of the cosmic shear measurement exhibit the highest signal-to-noise ratio \cite{Euclid:2019}. Their inclusion will be of the utmost importance to distinguish between competing gravity models with distinct nonlinear behaviors. Nontrivial effects that should be taken into account include nonlinear structure formation \cite{Takada:2003}, baryonic effects on the matter distribution \cite{Schneider:2019xpf}, intrinsic alignments \cite{Arico:2023} and the impact of massive neutrinos on structure formation \cite{Hamann:2012, Audren:2012}. For beyond-$\Lambda$CDM cosmologies, which include extensions of GR, the screening mechanism \cite{Brax:2013, Burrage:2017qrf, Koyama:2018som,Vardanyan:2023jkm} is an additional important nonlinear effect. Screening is a key phenomenological aspect of all viable Modified Gravity (MG) theories, allowing recovery of GR on small scales. At small scales, notably within our Solar System, GR exhibits highly accurate observational consistency (see Ref.~\cite{Will:2014kxa} for a review).

There is a plethora of non-standard cosmologies (for reviews, see e.g. Refs.~\cite{Clifton:2011jh, Joyce:2016vqv}), and a few of them have been tested with real photometric data (see, for example, Refs.~\cite{Liu:2016xes,KiDS:2020ghu, DES:2021zdr, DES:2022ccp}) and simulations \cite{Barreira:2016ias,Li:2017xdi,Hassani:2019lmy, Harnois-Deraps:2022bie}, or have been subject to forecasts for Stage-IV weak lensing surveys \cite{Schmidt:2008hc, Euclid:2023tqw,SpurioMancini:2023mpt}. An alternative approach to studying a broader class of MG models involves Horndeski theories \cite{Horndeski:1974wa}, the corresponding $\alpha$-parametrisation \cite{Bellini:2014fua,Lombriser:2018olq,Frusciante:2019xia} or direct parametrisation of the linear relation between matter and the gravitational potential (see, for example, Ref.~\cite{Zucca:2019xhg}). Typically, in such parameterised analyses, screening has either been neglected \cite{Srinivasan:2021gib, Srinivasan:2023qsu} or accounted for through conservative scale cuts \cite{Joudaki:2017zdt, Pogosian:2021mcs, DES:2022ccp, Wang:2023}.  

To our knowledge, attempts to take the screening effects into account have been made in Refs.~\cite{Alonso:2016suf, Reischke:2018ooh, SpurioMancini:2018apc, SpurioMancini:2019}. Therein, screening was modeled as an exponential cut-off in the $\alpha$-parameters with a characteristic screening scale at which the screening mechanism becomes effective. This scale was either varied as a free parameter or fixed to the value motivated by simulations \cite{Barreira:2013eea}. However, the performance of this approximation across various MG scenarios remains to be tested, in order to understand if the corresponding accuracy suffices for Stage-IV surveys.

In Ref.~\cite{Bose:2022}, we introduced an approximate phenomenological model based on the error function to model the screening effects within the halo model reaction framework~\cite{Cataneo:2018otr}. Various extended scenarios were shown to be in good agreement (within a few percent) with the exact solution for the screened regime. In the same paper, we described the minimal extension model, where the background is assumed to be flat $\Lambda$CDM, and the deviation of linear growth from GR is parameterised by the growth index $\gamma$ ~\cite{Peebles:1980,Linder:2005, Linder:2007}. The growth index is defined via the growth rate, $f = {\rm d} \ln D / {\rm d} \ln a$,  as an exponent of the time-dependent matter density relative to critical:
\begin{equation}
    f(z)=\Omega_{\rm m}^\gamma(z)\, .
\end{equation}
This phenomenological parameterisation was developed with galaxy clustering probes in mind and has been used to look for deviations from standard cosmology \cite{Alam:2015, Nguyen:2023fip, Moretti:2023drg}. In Ref.~\cite{Tereno:2011} it was demonstrated that $\gamma$ can be constrained with weak lensing probes as well. Constraining the growth index is one of the main objectives of Stage-IV surveys \cite{DESI:2016fyo, LSSTDarkEnergyScience:2018jkl, Euclid:2019}. We should note that this parametrisation assumes that the linear growth factor and rate are scale-independent which is not the case for some modified gravity models, such as the well studied $f(R)$ model of Ref.~\cite{Hu:2007nk}. To account for such scale dependencies one would need to extend this parametrisation as done in Ref.~\cite{Mirzatuny:2019dux}. We will not consider this here and restrict ourselves to the scale-independent case.

In this paper, we investigate whether we could provide constraints on the minimal model-independent parametrisation of MG, including nonlinear scales with forthcoming Stage-IV cosmic shear data. We present an emulator, trained with \texttt{cosmopower}\footnote{\href{https://github.com/alessiospuriomancini/cosmopower/tree/main}{github.com/alessiospuriomancini/cosmopower}} on power spectra generated using the halo model reaction code \texttt{ReACT}\footnote{\href{https://github.com/nebblu/ACTio-ReACTio}{github.com/nebblu/ACTio-ReACTio}}, which will be used in our analyses. We have made a public repository for this and related emulators \footnote{\href{https://github.com/nebblu/ReACT-emus/tree/main/emulators}{github.com/nebblu/ReACT-emus/emulators}}. 

In Section~\ref{sec:model}, we introduce the modelling of cosmic shear and relevant nonlinear effects. In Section~\ref{sec:setup}, we present our analysis setup. In Section~\ref{sec:results}, we study how well the nonlinear extension of the growth index performs on mock data in GR and an example MG model. Then, we investigate the importance of including the screening scale, the degeneracy between the screening scale and baryonic feedback, and the effects from massive neutrinos. Finally, we summarise our results and outline the next necessary steps, paving the way to an optimal model-independent parametrisation on all scales.

\section{Cosmic Shear Modelling}
\label{sec:model}
In this section, we describe the modelling of our observable, the cosmic shear power spectrum, detailing its main components, the MG scenarios under consideration, and their signatures on the weak lensing power spectrum.

\subsection{Cosmic Shear}

For a generic theory of gravity, the cosmic shear power spectrum can be derived from the lensing potential, also known as the Weyl potential. The Weyl potential is intricately linked to the underlying matter density fluctuations via a corresponding Poisson equation. Therefore, we can obtain the cosmic shear power spectrum by integrating the nonlinear matter power spectrum along the line-of-sight under the Limber approximation \cite{Reischke:2018ooh}:
\begin{equation}
\label{eq:Cell}
     C^{\gamma \gamma}_{ij} (\ell) = \int_{z_{\rm min}=0}^{z_{\rm max}} \mathrm{d}z \frac{W^{\gamma}_i(z)W^{\gamma}_j(z) }{H(z)\chi^2(z)} \Sigma^2[k_\ell(z), z] P_{\rm NL}[k_\ell(z), z] \, ,
\end{equation}
where we set the speed of light to $c=1$, 
$k_\ell = (\ell+1/2)/\chi(z)$, $\chi(z)$ signifies the radial comoving distance from an observer at $z=0$ to an object at redshift $z$, $H(a)$ represents the Hubble function, $\Sigma(k, z)$ denotes the modification to the Poisson equation for the Weyl potential in MG theories, and $z_{\rm max}$ stands for the maximum redshift of the source distribution in a survey. The weighting functions $W^\gamma_i$ are defined as:
\begin{equation}
\label{eq:Wi}
   W^{\gamma}_{i} (z) = \frac{3}{2} H_0^2 \Omega_{\rm m} \chi(z) (1+z) \int_{z_{\rm min}=0}^{z_{\rm max}} \mathrm{d}z' n_i(z') \left[1-\frac{\chi(z)}{\chi(z')} \right] \, ,
\end{equation}
where $\Omega_{\rm m}=\Omega_{\rm m}(z=0)$ denotes the total matter density fraction today, $H_0$ represents the Hubble constant in $\rm{Mpc}^{-1}$, and $n_i$ is the redshift distribution for bin $i$. The observed lensing signal $C^{\rm WL}_{ij}(\ell)$ is contaminated by the intrinsic alignment contribution (IA, see Ref.~\cite{Lamman:2023} for a recent review). IA represent correlations in the orientation between galaxies which are not caused by lensing but rather by the same gravitational field in which these galaxies evolve. We model this using the redshift-dependent nonlinear alignment (zNLA) model \cite{Bridle:2007ft,Mandelbaum:2009ck}. Additionally, we account for the shape-noise contribution due to the intrinsic ellipticity field (unlensed). These effects introduce the following additional contribution to the signal:
\begin{equation}
\label{eq:Cell_WL}
   C^{\rm WL}_{ij}(\ell) = C^{\gamma \gamma}_{ij} (\ell)+ C^{I \gamma}_{ij} (\ell)+C^{II}_{ij} (\ell)+\frac{\sigma_\epsilon^2}{\bar{n}} \delta^{\rm K}_{ij}\, ,
\end{equation}
where $\delta^{\rm K}_{ij}$ is the Kronecker's delta, $\sigma_\epsilon$ stands for the variance of the intrinsic ellipticity distribution, $\bar{n}= n/N_{\rm bins}$ with $n$ being number of galaxies per radians squared, and the superscript $I$ denotes the contribution from the IA with the following kernel:
\begin{equation}
\label{eq:WIi}
      W^{I}_{i} (k, z) = -A_{\rm IA} C_{\rm IA}(1+z)^{\eta_{\rm IA}}  \frac{\Omega_{\rm m}}{D(k, z)/D(k, 0)} n_i(z) H(z)\, , 
\end{equation}
where we set $C_{\rm IA}=0.0134$ (its conventional value, as it is degenerate with $A_{\rm IA}$), while $A_{\rm IA}$ and $\eta_{\rm IA}$ are additional nuisance parameters in our modelling. $D(k, z)$ denotes the growth factor\footnote{This is found by solving the linearised growth equation in the cosmological scenario of interest, which we compute using \texttt{MGrowth}:  \href{https://github.com/MariaTsedrik/MGrowth}{github.com/MariaTsedrik/MGrowth}.} -- the general definition for  MG theories includes a scale dependence, but in this work we only study models with scale-independent linear growth. In general, IA occurs on small astrophysical scales, where we assume that all modifications of gravity are screened \cite{Frusciante:2019xia}. The zNLA model allows for amplitude and redshift dependence of the IA signal from the gravity model through $D(k, z)$ in Eq.~\ref{eq:WIi} and $P_{\rm NL}$ in Eq.~\ref{eq:Cell}. Unlike other IA models where tidal physics is taken into account and perturbatively re-derived for MG scenarios, this does not affect NLA \cite{DES:2022ccp}. While there is still debate on whether or not zNLA captures all IA effects sufficiently well for Stage-IV surveys \cite{Leonard:2024nnw}, we leave a detailed exploration of more complex models for future work.

The main component from Eq. \ref{eq:Cell} that we model and modify in our analysis is the nonlinear power spectrum:
\begin{equation}
\label{eq:PNL}
   P_{\rm NL}(k, z) =  P^{\mathrm{MG}+\nu}_{\rm NL}(k,z) \times B^{\rm baryons}(k,z) = B^{\mathrm{MG}+\nu} \times B^{\rm baryons} \times P^{\Lambda \rm CDM}_{\rm NL}\, ,
\end{equation}
where the nonlinear $\Lambda$CDM power spectrum, $P^{\Lambda \rm CDM}_{\rm NL}$, is combined with the emulated boosts due to effects from baryonic feedback, $B^{\rm baryons}$, and MG with massive neutrinos, $B^{\mathrm{MG}+\nu}$. This prescription is described in detail in the following subsections. 

\subsection{Nonlinear Power Spectrum Modelling}
\label{subsec:Pnonlin}

In the halo model reaction framework\footnote{
The series of papers ``On the road to percent accuracy'' covers the developments of the framework including the basic foundation \cite{Cataneo:2018otr}, an emulator for the pseudo power spectrum \cite{Giblin:2019otr}, inclusion of massive neutrinos \cite{Cataneo:2019otr}, the first forecasts for an LSST-like survey and the \texttt{ReACT} code release \cite{Bose:2020otr}, a comparison with simulations and simulation-based emulators including baryonic contribution in \cite{Bose:2021otr}, and interactions between dark matter and dark energy \cite{Carrilho:2021otr}.} \cite{Cataneo:2018otr}, based on the halo-model approach (see Ref.~\cite{Asgari:2023} for a recent review), we can compute the nonlinear power spectrum in a modified theory of gravity including massive neutrinos as:
\begin{equation}
 P^{\mathrm{MG}+\nu}_{\rm NL}(k,z) = \mathcal{R}(k,z) P^{\rm pseudo}_{\rm NL}(k,z)\,,
 \label{eq:nonlinpk}
\end{equation}
where $P^{\rm pseudo}_{\rm NL}(k,z)$ is called the pseudo power spectrum. 
This is defined as a nonlinear power spectrum, evolved in a $\Lambda$CDM universe with adjusted initial conditions in order to match the linear clustering of the MG model of interest at the target redshifts, i.e., $P^{\rm pseudo}_{\rm L}(k,z_j)=P^{\rm MG}_{\rm L}(k, z_j)$. In the context of N-body simulations it can be understood as follows: initial conditions are generated from a power spectrum computed as $P^{\rm pseudo}_{\rm L}(k, z_{\rm ini})=D^2_{\Lambda\rm CDM}(z_{\rm ini})/D^2_{\Lambda\rm CDM}(z_j)~P_{\rm L}^{\rm MG}(k, z_j)$ while the evolution up to $z_j$ is done with standard gravity and $\Lambda$CDM expansion. In the context of \texttt{HMcode} it can be understood as using $D^2_{\Lambda\rm CDM}(z=0)/D^2_{\Lambda\rm CDM}(z_j)~P^{\rm MG}(k, z_j)$ as input linear power spectrum at redshift $z=0$. Since in our case the re-scaling of the initial conditions is scale-independent (as we are dealing with MG theories with scale-independent linear growth), one can also compute a pseudo power spectrum at the target redshift simply by re-scaling the primordial amplitude as $A^{\rm pseudo}_s =  D^2_{\rm MG}(z_j)/D^2_{\Lambda \rm CDM}(z_j) A_s$. Instead of utilising a $\Lambda$CDM nonlinear power spectrum with the same cosmological parameters and modelling the reaction as an expected ratio $P^{\mathrm{MG}+\nu}_{\rm NL}/P^{\Lambda \mathrm{CDM}}_{\rm NL}$ with halo model, we opt for the pseudo cosmology, which implies the halo mass functions of both MG and pseudo cosmologies are similar. As a consequence, the transition between 2- and 1-halo terms becomes smoother. For example, in Fig.~5 and~7 of Ref.~\cite{Cataneo:2018otr}, when comparing $P^{\mathrm{MG}}_{\rm NL}/P^{\Lambda \mathrm{CDM}}_{\rm NL}$ and $P^{\mathrm{MG}}_{\rm NL}/P^{\rm pseudo}_{\rm NL}$, the latter clearly appears as a smoother function.

We compute the pseudo power spectrum using \texttt{HMCode} \cite{Mead:2015,Mead:2016a,Mead:2016b,Mead:2020}, in particular its 2020-version. 
Alternatively, the pseudo power spectrum can be computed with \texttt{HaloFit} \cite{Smith:2002dz} or for MG theories with scale-independent linear growth, $\Lambda$CDM-based emulators such as \texttt{EuclidEmulator2} \cite{Knabenhans:2018, Euclid:2020} or \texttt{BACCOemu} \cite{bacco:2021} can be used by adjusting the spectrum amplitude parameter, $A_s$ or $\sigma_8$, to match the modified cosmology.    

The reaction function $\mathcal{R}(k,z)$ contains all nonlinear corrections to the pseudo spectrum coming from modifications of gravity and massive neutrinos. We refer the reader to Refs.~\cite{Cataneo:2018otr, Bose:2020otr} for more details on how to model the reaction function. In a nutshell, similar to the halo-model approach, there are three distinct regimes in scale: linear (2-halo term), quasi-nonlinear (1-loop corrections and smoothing), and nonlinear (1-halo term). For each we need to follow a certain prescription within our framework. These three regimes are demonstrated clearly in the Poisson equations that connect the gravitational potential, $\Phi$, to the matter density fluctuations, $\delta$:
\begin{align}
-\left(\frac{k}{a H(a)}\right)^2\Phi_{\rm QNL} (\bfk,a)=&
\frac{3 \Omega_{\rm m}(a)}{2} \mu_{\rm L}(k,a) \,\delta_{\rm QNL}(\bfk,a) + S(\bfk,a) \, , \label{eq:poisson1} \\ 
-\left(\frac{k}{a H(a)}\right)^2\Phi_{\rm NL} (\bfk,a)=&
\frac{3 \Omega_{\rm m}(a)}{2} \mu_{\rm NL}(k, a) \,\delta_{\rm NL}(\bfk,a) \, ,
\label{eq:poisson2}
\end{align} 
where $\Omega_{\rm m}(a)=\Omega_{\rm{m}}a^{-3} H_0^2/H^2(a)$. The subscripts QNL and NL denote ``quasi-nonlinear'' and ``nonlinear'', respectively.

In the linear regime, the enhancement or suppression of structure formation relative to $\Lambda$CDM is controlled by $\mu_{\rm L}(k, a)$. In the quasi-nonlinear regime, the modifications at second and third orders are captured by a source term $S(\bfk,a)$ (see Eq.~2.8 in Ref.~\cite{Bose:2016}). Within the reaction function, two parameters guarantee a smooth transition between the 2- and 1-halo terms in the quasi-nonlinear regime: $\mathcal{E}$ and $k_\star$. The first parameter corresponds to the 1-halo power spectrum ratio in the modified and pseudo cosmologies at very large scales, tuning the similarities in their halo mass functions. The second parameter controls the rate of the transition, and for that parameter the one-loop corrections given by $S(\bfk,a)$ are essential. 

In the nonlinear regime, the modification to gravity is governed by $\mu_{\rm NL}(k, a)$, which should recover GR at very nonlinear scales $(k\gg 10 \, h/\mathrm{Mpc})$ in the screened MG theories $\mu_{\rm NL}\rightarrow 1$ or  $\mu_{\rm NL}=\mu_{\rm L}$ in the unscreened MG theories. This function is then considered when solving the gravitational collapse equation for the top-hat radius. From the solution of the gravitational collapse equation and Virial theorem, we derive the density at the collapse, virial scale factor and virial density, which are then used for computing the halo-mass function \cite{Sheth:1999mn, Sheth:2001dp} and halo density profile \cite{Navarro:1996gj} -- two essential properties of the 1-halo term power spectrum.

All in all, to fully describe $P^{\mathrm{MG}+\nu}$ we require the following information: the total neutrino mass $M_\nu$  (or equivalently the neutrino density parameter, related to the mass by  $M_\nu \approx 93.14 ~\Omega_\nu h^2$ eV), and 5 functions -- the expansion history $H(a)$, the modification of gravity on linear scales $\mu_{\rm L}(k, a)$, two functions $\gamma_2$ and $\gamma_3$ included in the source term $S(\bfk,a)$ that modifies quasi-nonlinear scales, and the modification of gravity in the nonlinear regime $\mu_{\rm NL}(k, a)$. In the subsequent subsection, we discuss these functions for the MG theories of interest.

\subsection{Modified Gravity Scenarios}
\label{sec:MGscenarios}

In our previous work \cite{Bose:2022}, we studied Horndeski theories employing the Effective Field Theory of Dark Energy (EFTofDE). Horndeski theories \cite{Horndeski:2024sjk} encompass the most general class of scalar-tensor theories of gravity in 4 dimensions that are Lorentz-covariant, ghost-free and have second-order equations of motion. However, in this work, we specifically focus on a popular example of scalar-tensor theories with Vainshtein screening \cite{Vainshtein:1972} -- the normal branch of DGP theories (nDGP, \cite{Dvali:2000}). We aim to recover its main features with the phenomenological growth index parameterisation \cite{Peebles:1980, Linder:2005, Linder:2007} extended to the nonlinear regime (see the ``minimal parametrisation" in Ref.~\cite{Bose:2022}).

Considering linear scalar perturbations on a Friedmann-Robertson-Walker metric, one can write the line element in the conformal Newtonian gauge and a spatially flat background as:
\begin{equation}
\mathrm{d}s^2 = -(1+2\Phi)\mathrm{d}t^2 + a^2(1-2\Psi) \mathrm{d}\bfx^2 \, ,
\end{equation}
with the Newtonian potential $\Phi$ and the intrinsic spatial curvature potential $\Psi$. We can then modify the linear Poisson equations neglecting the anisotropic shear as
\begin{align}
-\left(\frac{k}{a H(a)}\right)^2\Phi_{\rm L} (\bfk,a)=&
\frac{3 \Omega_{\rm m}(a)}{2} \mu_{\rm L}(k,a) \,\delta_{\rm L}(\bfk,a)  \, , \label{eq:poisson_lin} \\ 
 \Psi_{\rm L} =& \eta_{\rm L}(k,a) \Phi_{\rm L}\, ,
\label{eq:eta}
\end{align} 
where $\eta_{\rm L}$ is the slip parameter, and which simplifies to the standard GR case for $\mu_{\rm L}=\eta_{\rm L}=1$. We now obtain a Poisson equation for the Weyl potential, $\Psi_{\rm W} = (\Phi_{\rm L}+\Psi_{\rm L})/2 $ as 
\begin{equation}
-\left(\frac{k}{a H(a)}\right)^2 \Psi_{\rm W} (\bfk,a)=
\frac{3 \Omega_{\rm m}(a)}{2} \Sigma_{\rm L}(k,a) \,\delta_{\rm L}(\bfk,a) \, ,
\end{equation}
with $\Sigma_{\rm L}(k,a)=\mu_{\rm L}(1+\eta_{\rm L})/2$, which equals unity in the GR case. Generally, most MG theories exhibit a strong preference to a no-slip condition so that $\Phi_{\rm L}\approx \Psi_{\rm L}$ (see Fig.~2-4 in Ref.~\cite{Peirone:2017ywi}), hence $\Sigma_{\rm L}=\mu_{\rm L}$. We can further extend the equality of the potentials into the nonlinear regime, which implies $\Sigma(k, z) = \mu_{\rm NL}(k, z)$ in Eq.~\ref{eq:Cell}: forcing not only $P_{\rm NL}$ but also $\Sigma$ to converge to a GR-limit on small scales. However, for our theories of interest we have $\Sigma_{\rm L} =1$, i.e., lensing is not affected: for nDGP this is derived in Refs.~\cite{Koyama:2005, Koyama:2007} and for the $\gamma$-parametrisation we set it by hand. Therefore, in this work, in Eq.~\ref{eq:Cell} we set $\Sigma(k, z) = 1$ and modify $P_{\rm NL}$ only.  

\begin{figure}
    \centering
    \includegraphics[width=0.49\columnwidth]{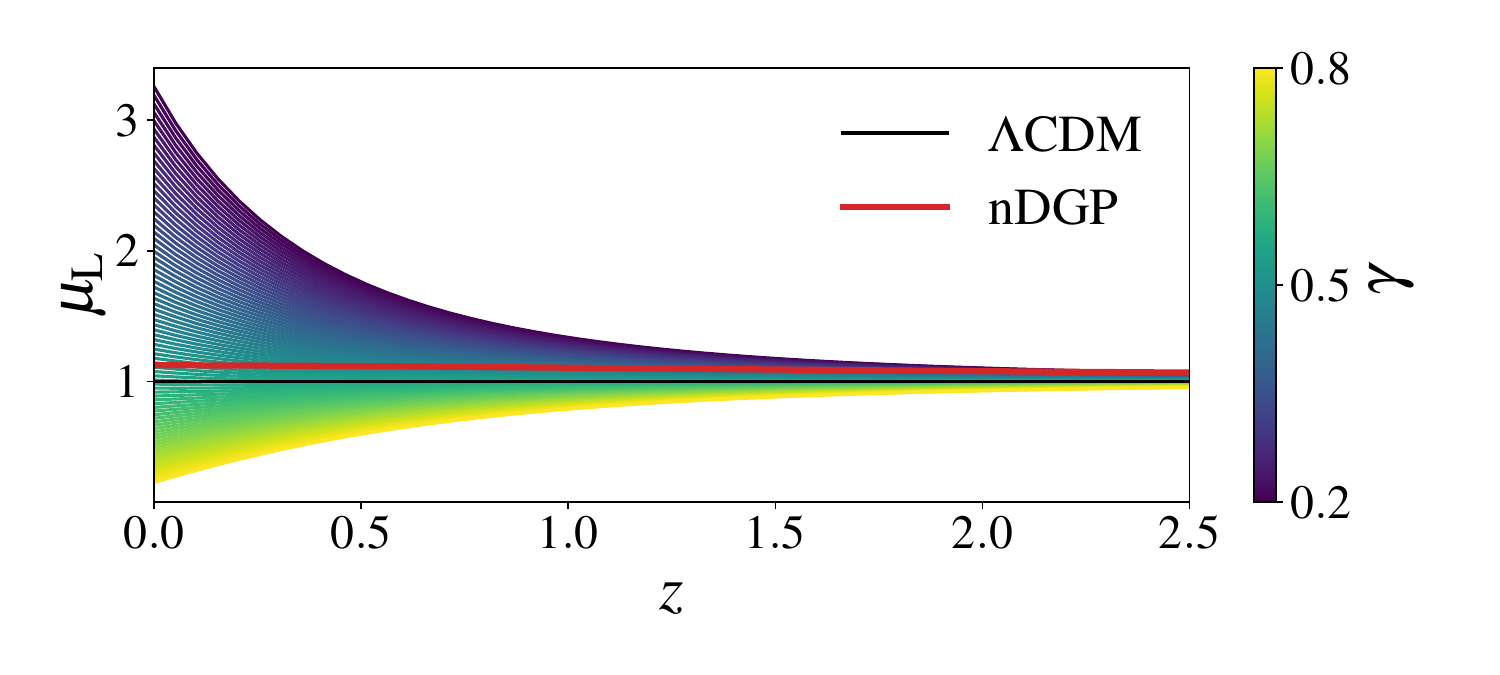}\includegraphics[width=0.49\columnwidth]{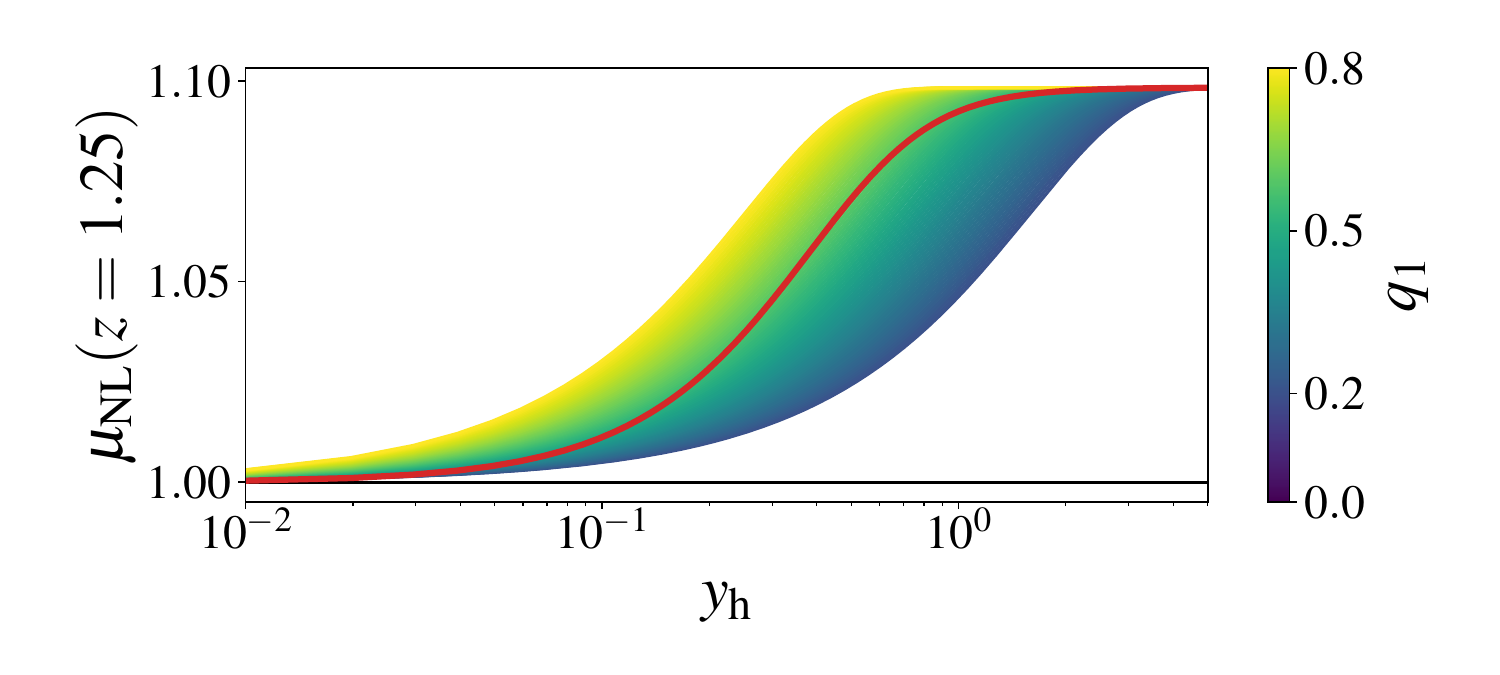}
    \includegraphics[width=\columnwidth]{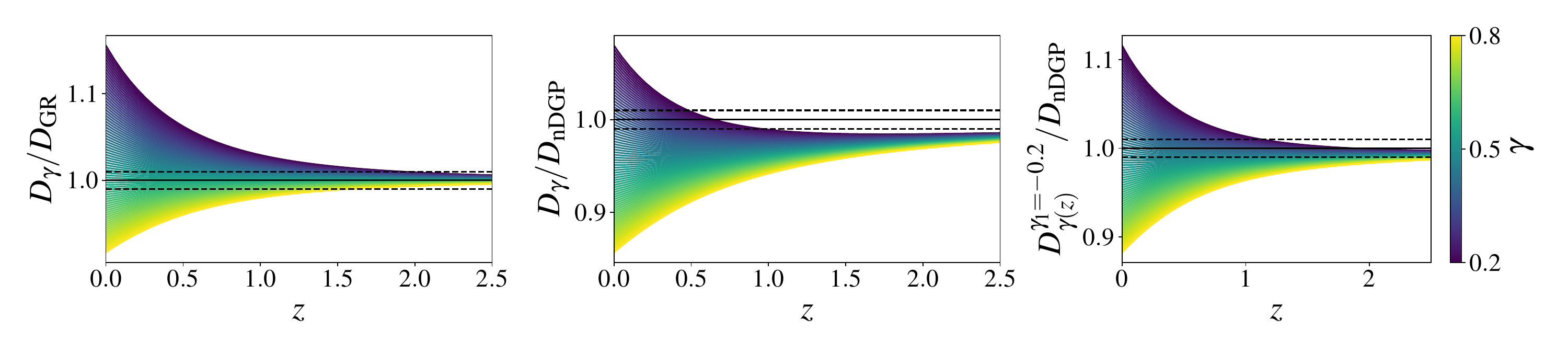}
    \caption{\textit{Upper panel}: left -- linear modification to the Poisson equation for nDGP gravity with strong modification of $\Omega_{rc}=0.25$ (red solid line), GR (black line) and different growth index values specified in the colorbar; right -- nonlinear modification to the Poisson equation for the same nDGP scenario, and growth index with $\gamma=0.4$ and various values of the screening scale $q_1$ at $z=1.25$ as a function of the normalised halo radius $y_{\rm h}$, defined in Eq.~\ref{eq:ydef}. \textit{Lower panel}: ratio between the growth factors computed for various values of $\gamma$; left -- with respect to the growth in GR, middle and right -- with respect to nDGP for time-independent and time-dependent growth index respectively. Dashed black lines denote the $1\%$ range. }
    \label{fig:MGrowths}
\end{figure}

Below is a summary of the functions required for the reaction calculations and our assumptions:
\begin{enumerate}
    \item The background expansion is set to be equal the standard cosmology: $H(a)= \\ H_0 \sqrt{\Omega_{\rm m}a^{-3}+(1-\Omega_{\rm m})}$.
    \item Modifications in the linear regime: 
    \begin{equation}
    \label{eq:mu_L}
        \mu_{\rm L}(a) = \begin{cases} 1+\frac{1}{3 \beta(a)} & \text{ for nDGP} \\
                                        \frac{2}{3} \Omega_{\rm m}^{\gamma-1}(a)\left[\Omega_{\rm m}^\gamma(a)+2-3\gamma+3(\gamma-1/2)\Omega_{\rm m}(a) \right] & \text{ for $\gamma=const$} \\
                                        \frac{2}{3} \Omega_{\rm m}^{\gamma-1}(a)\gamma_1(a-1/a)\log{\Omega_{\rm m}(a)}+\mu^{\gamma=const}_{\rm L}(a) & \text{ for $\gamma(a)$}
                                        \end{cases}
    \end{equation}
    with $\beta(a)$ defined as 
    \begin{equation}
    \label{eq:beta}
    \beta(a)=1+\frac{H(a)}{H_0}\frac{1}{\sqrt{\Omega_{rc}}}\left( 1+ \frac{\dot{H}(a)}{3H^2(a)}\right)\, ,
    \end{equation}
    where dot denotes a derivative with respect to the time coordinate and $\Omega_{rc}$ corresponds to the strength of modification in the nDGP cosmology ($\Omega_{rc}=0$ in GR). In addition to the standard (constant) growth index parameter, we consider a time-dependent version of the form $\gamma(a)=\gamma_0+\gamma_1(a+1/a-2)$ from the recent work of Ref.~\cite{Wen:2023}. We add this model to \texttt{MGCAMB}\footnote{\href{https://github.com/MariaTsedrik/MGCAMB}{github.com/MariaTsedrik/MGCAMB}}.
    The latter parametrisation was recently proposed because it captures the time-evolution of growth in Horndeski theories better than the constant growth index. We demonstrate this in the lower right panel of Fig.~\ref{fig:MGrowths}. Additionally, in the same figure, we illustrate the impact different choices of $\mu_{\rm L}$ have on the growth factor $D$ for the time-independent growth index. Indeed, for nDGP the time-dependent parametrisation of the growth index provides a better fit than the constant one. However, for the standard cosmology, the Linder gamma parametrisation remains excellent, whereas, for a scalar-tensor theory like nDGP, it does not. Further discussion on the correspondence between the constant growth index and DGP theories is provided in Appendix~\ref{sec:AppB}. In all models, we ensure a match with General Relativity at high redshifts. In other words, we introduce modifications of the standard cosmology only in the late universe, while the early universe remains $\Lambda$CDM.
    
    \item In the quasi-nonlinear regime for nDGP $\gamma_2$- and $\gamma_3$-functions within $S(\bfk, a)$ are specified in Ref.~\cite{Bose:2016}, while for the scale-independent growth index parametrisation we set them to zero. In Ref.~\cite{Bose:2022} it was demonstrated that the 1-loop corrections impact the reaction function very weakly for nDGP, less than $1\%$. However, this might become a consideration when scale-dependency in the linear growth is present.
    
    \item Modifications in the fully nonlinear regime:
     \begin{equation}
     \label{eq:muNL}
        \mu_{\rm NL}(a) = \begin{cases} \frac{2}{3\beta(a)} \frac{\sqrt{1+s^{3}} -1}{s^{3}}+1 & \text{ for nDGP} \\
                                        (\mu^{\gamma}_{\rm L}(a)-1)\mathrm{erf}(a \, y_{\rm h} 10^{q_1})+1 & \text{ for $\gamma$ parametrisations} \,.
                                        \end{cases}
    \end{equation}
    The expression for nDGP is derived in Ref.~\cite{Schmidt:2009} following the solution of a spherically symmetrical overdensity with 
    \begin{equation}
    s  = \left[ \frac{2 \Omega_{\rm m} (\delta+1) }{9 a^3 \beta(a)^2 \Omega_{rc}} \right]^{\frac{1}{3}},
    \end{equation}
    and $\delta$ being the nonlinear over-density given by 
\begin{equation}
    \delta = y_{\rm h}^{-3} (1+\delta_{\rm ini}) -1 \, , \label{eq:nldensity} 
\end{equation}
with $\delta_{\rm ini}$ being the initial over-density and 
\begin{equation}
    y_{\rm h} \equiv \frac{R_{\rm TH}/a}{R_i/a_{\rm ini}} \, , \label{eq:ydef}
\end{equation}
$R_{\rm TH}$ and $R_i$ being the physical halo top-hat radius at the target scale factor $a$ and at the initial scale factor $a_{\rm ini}$, respectively. The second expression is phenomenological and was demonstrated in Ref.~\cite{Bose:2022} to reproduce the Vainshtein screening behaviour within $1\%$ up to $k=5~h/\mathrm{Mpc}$ when compared against the exact solution. The parameter $q_1$ corresponds to the screening scale, and its impact is demonstrated in the right upper panel of Fig.~\ref{fig:MGrowths}. From this figure, we see that the higher $q_1$ the deeper into the halo the transition from modification to no modification happens. The opposite is also true, the lower $q_1$ the farther outside the halo $\mu_{\rm NL}=1$. The GR-limit is fully recovered at all scales when $q_1 \rightarrow -\infty$. Values of $q_1 \gtrsim 2$ correspond to  a screening scale deep within a halo, indicating that we will observe no effect of screening in the nonlinear power spectrum at the scales of our interest.
\end{enumerate}

\subsection{Baryonic Effects}
\label{sec:baryonic_effect}

The last missing component from Eq.~\ref{eq:nonlinpk} is the boost from the baryonic feedback. It is well-known that astrophysical processes significantly impact the matter power spectrum in the nonlinear regime. For instance, active galactic nuclei (AGN), supernovae, and stellar winds repulse matter from clustered centralised clumps into intergalactic medium, causing a suppression of the observed structures on small scales. On even smaller astrophysical scales, the observed structures get boosted by gas cooling and star formation. The effects of baryonic processes have been extensively studied in various hydrodynamical simulations, including  COSMO-OWLS \cite{Brun:2013}, BAHAMAS \cite{McCarthy:2016}, Illustris-TNG \cite{Springel:2017}, and more recent FLAMINGO \cite{Schaye:2023jqv}, among others. However, these effects are sensitive to the sub-grid physics and can vary significantly between different simulations (as depicted in Fig.~9 of Ref.~\cite{Schneider:2018}).

Instead of running computationally expensive hydro-simulations and fine-tuning them to match observations, the impact of baryons can be modelled using a baryonification model \cite{Schneider:2015,Schneider:2018}. In this approach, halo profiles, the outputs of a dark-matter-only simulations, are modified by slightly displacing particles around halo centres in a spherically symmetric way. This displacement is characterised by 7 baryonic parameters: ${\log_{10} M_c, \, \theta_{\rm ej}, \, \mu, \gamma, \, \delta, \, \eta_\delta, \, \eta}$ (the first five describe the gas distribution, while the last two are related to the stellar abundances), along with one cosmological parameter, the baryonic fraction $f_{\rm b}=\Omega_{\rm b}/\Omega_{\rm cdm+b}$. In our analysis we use the baryonic emulator \texttt{BCEMU} \cite{Giri:2021}, which accurately reproduces the power spectra of state-of-the art hydrodynamical simulations even with a reduced number of parameters. This emulator has been employed to analyse cosmic shear data in Ref.~\cite{Schneider:2021}, and provides a robust and flexible tool for incorporating baryonic effects into our analyses. In Section~\ref{sec:GQbar} we discuss the cosmology dependence in the context of MG theories for the baryonification approach.

 The importance of baryonic feedback becomes evident on scales comparable to the size of a halo, which coincides with the scales where screening is crucial for MG theories. In this work, we aim to investigate whether an independent parametrisation of the screening scale can be decoupled from the effects of baryons. To achieve this, when exploring the generalised parametrisation in Sections~\ref{sec:GQbar} and \ref{sec:DGPall}, we only vary $\log_{10} M_c$ (similarly to the DES Y3 shear analysis in Ref.~\cite{DES:2022}), while keeping the others fixed at their fiducial values. Exploring degeneracies with the full set of baryonic parameters will be explored in future work. The $\log_{10} M_c$ parameter controls the slope of the gas density distribution: smaller haloes with masses less than $M_c$ have shallower profiles. Both the screening and baryonic feedback effects display weak cosmological dependence and result in the suppression of growth on approximately the same scales. Already from this fact alone we can highlight the urgency of accurately measuring baryonic parameters. 
 
Fortunately, there exists a well-established connection between the baryonic effects on the power spectrum and the gas and stellar fractions in haloes \cite{Schneider:2015}. This suggests that additional CMB and/or X-ray measurements of galaxy groups and clusters can help constrain these parameters or impose physically motivated priors.
For instance, in Ref.~\cite{Grandis:2023qwx} authors measured $\log_{10}{M_c} = 14.53 \pm 0.20$ with a compilation of Bayesian population studies of galaxy groups and clusters and with cluster gas density profiles derived from deep, high-resolution X-ray observations. While Ref.~\cite{Schneider:2021wds} found from the combination of the cosmic shear data from KiDS-1000 with the gas profiles from kinematic Sunyaev–Zeldovich (kSZ) observations and X-ray data constrained $\log_{10}{M_c} \approx 13.2 \pm 0.4$. Recently, Ref.~\cite{DES:2024iny} found a significantly stronger baryonic feedback suppression than in previous studies using the DES Y3 cosmic shear and CMASS + ACT DR5 kSZ observations with $\log_{10}{M_c} = 13.22^{+0.42}_{-0.29}$. In these three works all 7 parameters of the model were varied. Interestingly, it has been demonstrated in Ref.~\cite{Schneider:2019xpf} that with the addition of mock gas fractions from eROSITA a Euclid-like setup finds $\sigma_{\log_{10}{M_c}}/\log_{10}{M_c} \sim \mathcal{O}(0.001)$. This implies that future surveys can decrease baryonification model errors by more than an order of magnitude.

Alternative approaches to mitigate the baryonic effects include a) a slightly different model of baryonification called Baryon Correction Model \cite{Arico:2019,Arico:2020} that was applied to DES Y3 data \cite{DES:2022,Arico:2023}, b) halo-model \cite{Semboloni:2011, Mead:2020}, and c) Principal Component Analysis (PCA) \cite{Eifler:2014, Mohammed:2017, Huang:2018}.

\section{Analysis Setup}
\label{sec:setup}

The survey specifications in our pipeline are chosen to mimic Stage-IV specifications: sky fraction $f_{\rm sky}= 0.4$, number density of galaxies per arcminute squared  $n= 30$ arcmin$^{-2}$, and per-component dispersion in the intrinsic galaxy ellipticities $\sigma_\epsilon = 0.3$. We take 10 equipopulated tomographic bins between redshifts $z \in [0.001, 2.5]$. The photometric redshift distribution of galaxies is following $n_i(z) \propto (z/z_0)^2 \exp{-\left( z/z_0\right)^{3/2}}$ with $z_0=0.9/\sqrt{2}$ \cite{Joachimi:2009} . The uncertainty on the photometric redshift is modeled as the sum of two Gaussian distributions: one for the well determined photometric redshifts and another for the outliers (we follow the prescription in Section 3.3.1 of Ref.~\cite{Euclid:2019}).

As input we use synthetic data, namely $C_\ell$ computed according to Eq.~\ref{eq:Cell}, with the fiducial cosmology in two scenarios: 
\begin{enumerate}
    \item $\Lambda$CDM scenario: nonlinear power spectrum with \texttt{HMcode2020} in which linear input is computed with \texttt{BACCOemu};
    \item nDGP scenario with strong deviation from the standard cosmology $\Omega_{rc}=0.25$: nonlinear power spectrum computed with reaction from \texttt{ReACT} and pseudo-power spectrum from \texttt{HMcode2020}.
\end{enumerate}
We use these scenarios to validate the pipeline and then test the performance of the growth index parametrisation. In total, we emulate 3 models using \texttt{cosmopower} \cite{SpurioMancini:2019}: nDGP, $\gamma+q_1$ and $\gamma(z)+q_1$ in the ranges specified in Table~\ref{tab:emulators} and the accuracy of emulators provided in the corresponding code repository\footnote{\href{https://github.com/nebblu/ReACT-emus/tree/main/emulators/Accuracy\%20Plots}{nebblu/ReACT-emus/emulators/Accuracy Plots}}. 
We compute our models in the wavenumber range of $k\in[0.01, 5] \, h/\mathrm{Mpc}$. For smaller $k$-values we extrapolate the boost to be constant and equal to its value at $k=0.01\, h/\mathrm{Mpc}$, while for larger $k$-values we extrapolate the boost as a power law.

The likelihood is computed as (see Appendix~B.2 of Ref.~\cite{Audren:2012}):
\begin{equation}
    \mathcal{L}=-\frac{1}{2} \sum_{\ell=10}^{\ell_{\rm max}} (2\ell+1) f_{\rm sky} \left(\frac{d_\ell^{\rm mix}}{d_\ell^{\rm model}} + \ln{\frac{d_\ell^{\rm model}}{d_\ell^{\rm data}}}-N_{\rm bins} \right)
\end{equation}
where $N_{\rm bins}$ is the number of photometric bins and $d_\ell$ denote the determinants of $C_\ell^{ij}$: the mock data and model are both computed with Eq.~\ref{eq:Cell_WL}, and $d_\ell^{\rm mix}$ is defined as
\begin{equation}
d_\ell^{\rm mix} = \sum_{k=1}^{N_{\rm bins}} \det \left[ \begin{cases}
    C^{\rm model}_{ij}(\ell)& \text{if } j\neq k\\
    C^{\rm data}_{ij}(\ell)&   \text{if } j= k
\end{cases}\right]\, .
\end{equation}
The modelling and likelihood function we use were validated in Ref.~\cite{Euclid:2023pxu}. The priors are given in Table~\ref{tab:model}, and the fiducial values for the fixed baryonic parameters are: $\theta_{\rm ej}=4.235$, $\mu= 0.93$, $\gamma=2.25$, $\delta = 6.4$, $\eta_\delta=0.14$, $\eta=0.15$.

\begin{table}[ht!]
\footnotesize
  \centering
  \begin{tabular}{l|lcc}
    Effect & Parameter & Prior & Fiducial \\
    \hline
    \multirow{5}{*}{Cosmological} & $\omega_{c}$ & $\mathcal{U}(0.11, 0.13)$ & 0.12 \\
    & $\omega_{b}$ & $\mathcal{N}(0.02268, 0.00038)$ & 0.02268 \\
    & $h$ & $\mathcal{U}(0.63755, 0.7305)$ & 0.68\\
    & $\log{\left(10^{10}~A_s \right)}$ & $\mathcal{U}(2.7081, 3.2958)$ & 3.044\\
    & $n_s$ & $\mathcal{N}(0.97, 0.004)$ & 0.97 \\
    \hline
    \multirow{2}{*}{Intrinsic Alignments} & $A^{\rm IA}$ & $\mathcal{U}(0, 12.1)$ & 1.72 \\
    & $\eta^{\rm IA}$ & $\mathcal{U}(-7, 6.17)$ & -0.41 \\
    \hline
    \multirow{4}{*}{Modified Gravity} & $\log_{10} \Omega_{rc}$ & $\mathcal{U}(-3, 2)$ & 0.25\\
    & $\gamma, \gamma_0$ & $\mathcal{U}(0, 1)$ & -\\
    & $\gamma_1$ & $\mathcal{U}(-0.7, 0.7)$ & -\\
    & $q_1$ & $\mathcal{U}(-2, 2)$ & -\\
    \hline
    \multirow{1}{*}{Baryonic} & $\log_{10} M_c$ & $\mathcal{U}(11.4, 14.6)$ & 13.32\\
    \hline
    \multirow{1}{*}{Massive Neutrinos} & $M_\nu$ & $\mathcal{U}(0, 0.1576)$ & 0.06
  \end{tabular}
  \caption{Sampling parameters, prior ranges and fiducial values in the model and mock data respectively. The spectral index $n_s$ has a tight Planck priors and the baryonic density component has a BBN-prior, both motivated by weak constraints on these parameters from cosmic shear data \cite{Hall:2021qjk}. 
  }
  \label{tab:model}
\end{table}
\begin{table}[ht!]
\footnotesize
  \centering
  \begin{tabular} { l c c c c c c c c c c c}
     & $\Omega_{\rm m}$  & $\Omega_{\rm b}$ & $\Omega_{\nu}$  & $h$ & $n_s$ & $A_s$ & $\gamma$, $\gamma_0$ &  $\gamma_1$&$q_1$ & $\log_{10} \Omega_{rc}$ & $z$\\
    \hline
lower & 0.2899 & 0.04044 & 0. &    0.629   &  0.9432&1.5 $\times 10^{-9}$& 0. & -0.7&-2 & -3&0.\\
upper & 0.3392 & 0.05686 & 0.00317 & 0.731&  0.9862&2.7 $\times 10^{-9}$& 1. & 0.7& 3 & 2. &2.4\\
    \hline
  \end{tabular}
  \caption{Prior ranges for the emulators.}
  \label{tab:emulators}
\end{table}
The posterior distribution is sampled with \texttt{Nautilus} \cite{Lange:2023} for $3\times10^3$ live points. This sampler is based on the importance nested sampling (INS) technique combined with deep learning via neural network regression. Similarly to other nested sampling-based samplers, \texttt{Nautilus} tends to under-predict the size of the posterior contours when compared to much slower but more accurate samplers. This is a known feature (see, for example, Appendix D of Ref.~\cite{Kilo-DegreeSurvey:2023gfr}) and not relevant for the purpose of this work. We varied the number of live points to guarantee the convergence and to control the differences in posterior-volumes when comparing between models with different number of parameters.

\section{Results}
\label{sec:results}

We first validate our models on synthetic data in $\Lambda$CDM, i.e. data with GR as the model of gravity. Then we test how the index growth models -- with and without screening -- perform when the true cosmology is nDGP, i.e., with enhancement of structure on large scales and recovering of GR on small-scales via Vainshtein screening. After that, we produce synthetic data with the screened $\gamma+q_1$ model, and study degeneracies between the screening scale and baryonic feedback, and massive neutrinos. Finally, we compare the performance of the exact nDGP model versus the model-independent growth index approach on a mock data vector assuming nDGP gravity, baryonic feedback and massive neutrinos.    

\subsection{Validation with GR}
\label{sec:GRvalisation}

\begin{figure}
    \centering
    \includegraphics[width=0.6\columnwidth]{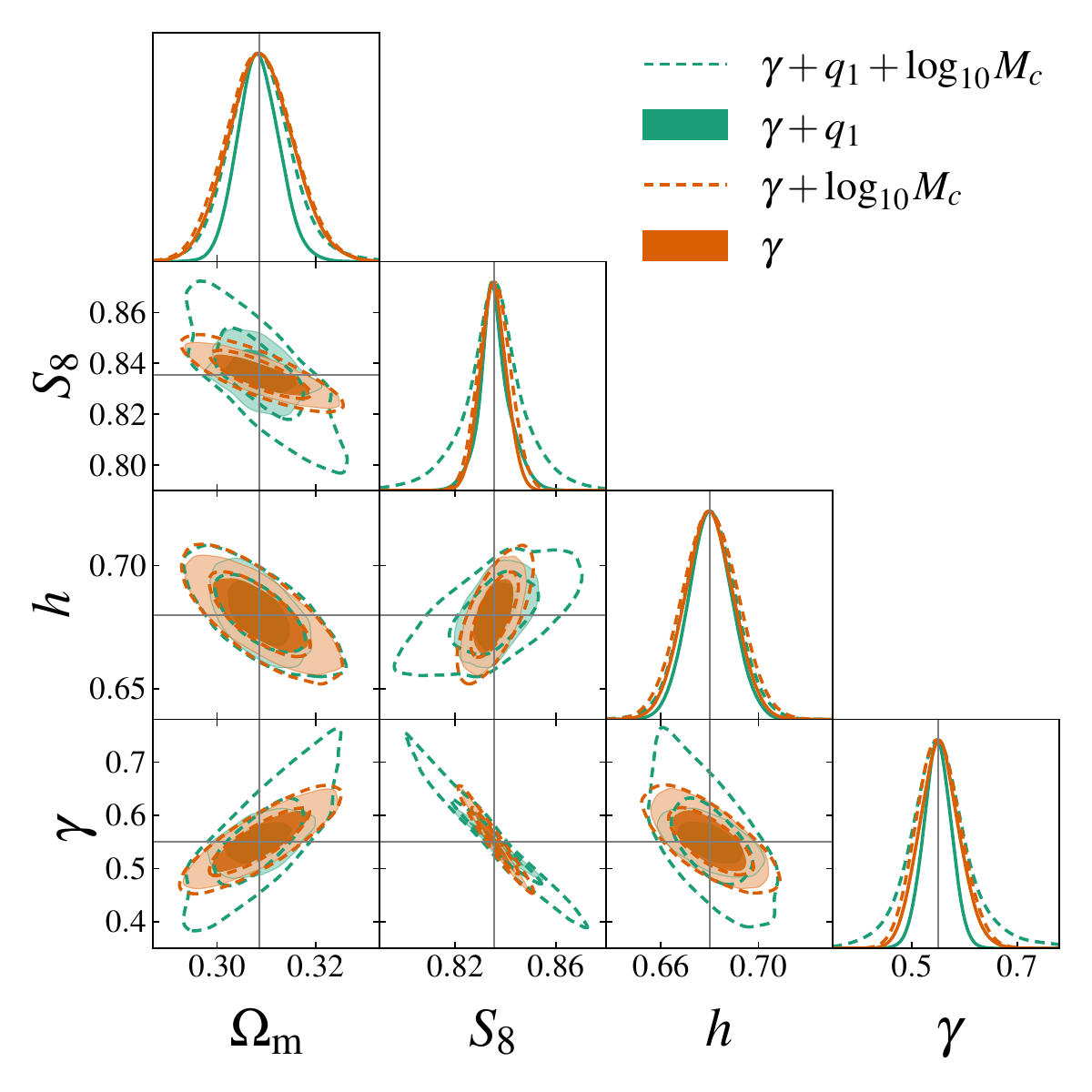}
    \caption{Marginalised posterior distribution for the cosmological
      parameters for 
      the two model choices,
      as detailed in the legend. We fit the cosmic shear power spectra with
      $\ell_{\rm max}=3000$. Solid lines and filled contours correspond to $B^{\rm baryons}=1$ from Eq.~\ref{eq:PNL} (for both -- data and models), while dashed lines and empty contours correspond to the baryonic feedback contribution with varying $M_c$ and fixed values of the other baryonic parameters as detailed in the main text. Grey lines mark the
      true values of the synthetic data.}
    \label{fig:LCDM-cosmo}
\end{figure}
\begin{figure}
    \centering
    \includegraphics[width=0.53\columnwidth]{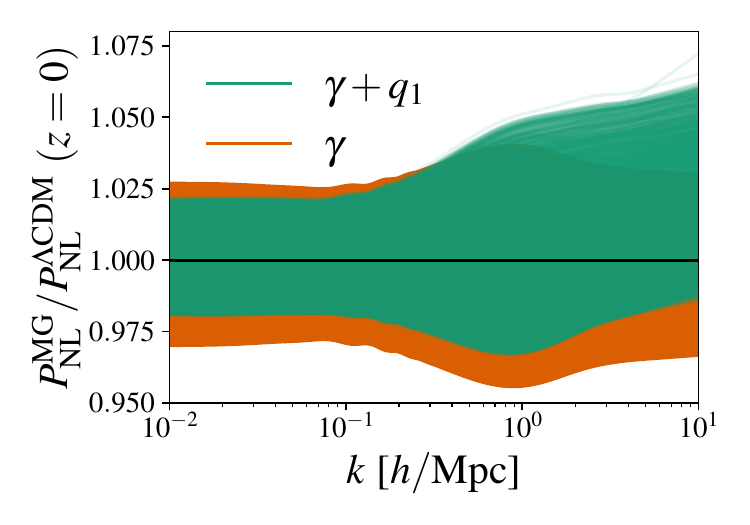}
    \includegraphics[width=0.46\columnwidth]{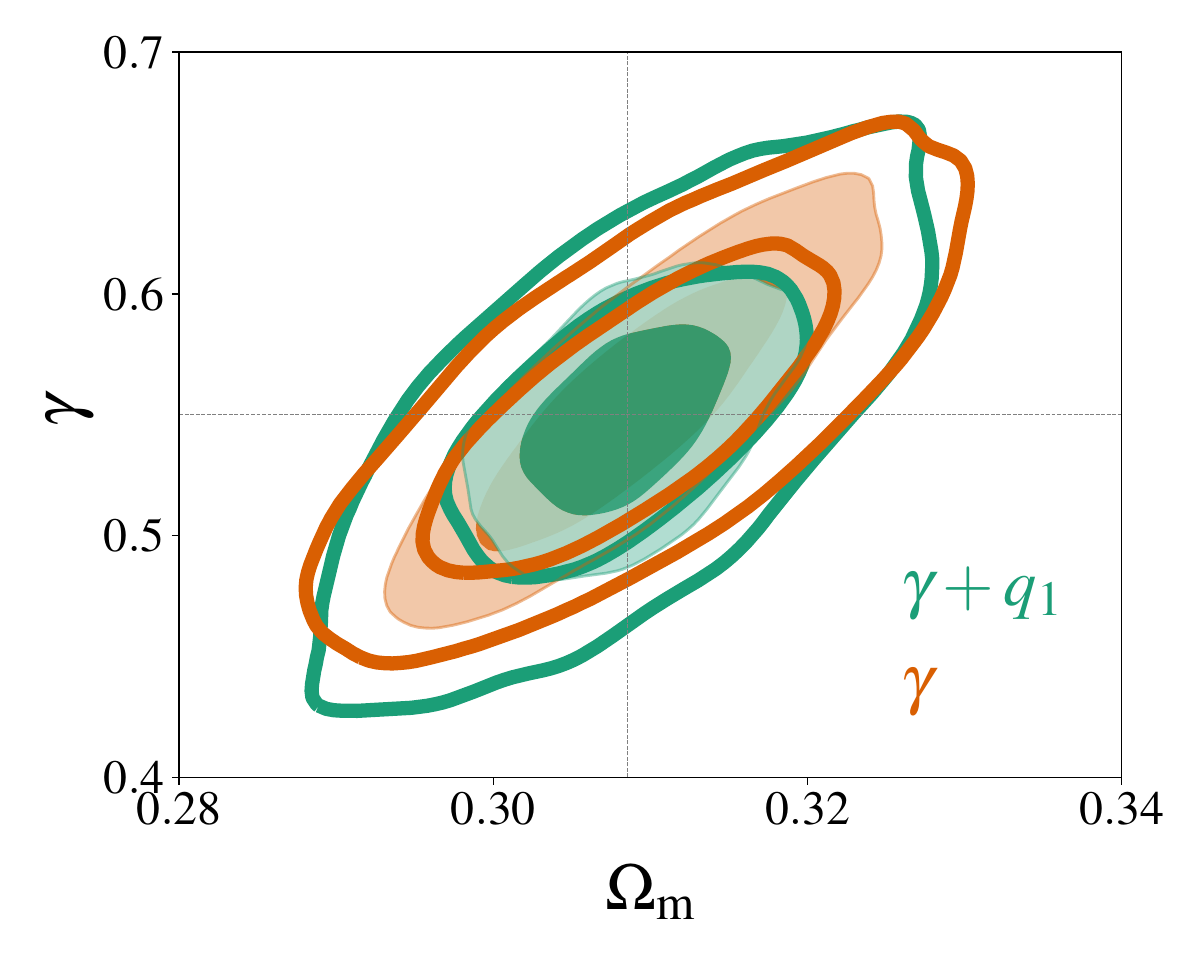}
    \caption{\textit{Left panel}: variation in the ratio $P^{\rm MG}_{\rm NL}/P^{\Lambda \rm CDM}_{\rm NL}$ at redshift zero for the screened, $\gamma+q_1$, and unscreened, $\gamma$, models drawn from the posterior distributions in Fig.~\ref{fig:LCDM-cosmo} within the $1\sigma$ deviation around the posterior maximum for $\gamma$. \textit{Right panel}: marginalised posterior distribution in $\Omega_{\rm m}-\gamma$ for the same models on the mock GR data with $\ell_{\rm max}=3000$ (filled contours) and $\ell_{\rm max}=1000$ (unfilled contours).}
    \label{fig:LCDMboost}
\end{figure}

In order to validate our likelihood pipeline, we perform MCMC analyses on mock data in the standard cosmology (without massive neutrinos). For the scale-cut we take $\ell_{\rm max}=3000$, which is considered a middle value between ``pessimistic'' ($\ell_{\rm max}=1500$) and ``optimistic'' ($\ell_{\rm max}=5000$) scenarios for Stage-IV surveys \cite{Euclid:2019}. We test two models -- gamma without screening via a pseudo-power spectrum ($\gamma$) and gamma with screening ($\gamma+q_1$). In Fig.~\ref{fig:LCDM-cosmo} we show posterior distributions for the main parameters of interest in weak lensing as well as the expansion rate today, $h$. The latter is added to demonstrate its anti-correlation with the extended parameter $\gamma$, while the $h$ degeneracy with $\omega_b$ and $n_s$ is broken by informative priors on the latter two parameters. The full posterior distribution is demonstrated in Fig.~\ref{fig:LCDM_valid_full} of Appendix~\ref{sec:GRdiscussion} together with a detailed discussion on the various degeneracies and the parameters controlling the amplitude\footnote{For the amplitude of the power spectrum we present both -- the sampled parameter of the primordial amplitude $A_s$, and its derived late-Universe counterpart $\sigma_8=D_\gamma(z=0)/D_{\Lambda \rm CDM}(z=0) \, \sigma^{\Lambda \rm CDM}_8$, with $\sigma_8$ being the r.m.s. density variation when smoothed with a tophat-filter of radius $8~\mathrm{Mpc}/h$, and $S_8=\sigma_8\sqrt{\Omega_{\rm m}/0.3}$. We also show the derived matter density parameter $\Omega_{\rm m}=(\omega_c+\omega_b)/h^2$.}.

In Fig.~\ref{fig:LCDM-cosmo}, both models recover all cosmological parameters correctly, and the value of gamma is $ \sim 0.55$ as expected, with uncertainties $\sigma_\gamma = 0.03$ for $\gamma+q_1$ and $\sigma_\gamma = 0.04$ for $\gamma$ without screening. The fact that a model with more parameters results in tighter constraints is explained by the functional form of $\mu_{\rm NL}$ (see Eq.~\ref{eq:muNL}): any deviation from $\gamma \sim 0.55$ or $\mu_{\rm L} = 1$ opens the possibility of constraining $q_1$, which leads to a worse fit or smaller likelihood due to nonlinear signatures that are very different from the GR predictions. We illustrate this in the left panel of Fig.~\ref{fig:LCDMboost}, where we plot the power spectra ratio $P^{\rm MG}_{\rm NL}/P^{\Lambda \rm CDM}_{\rm NL}$ at $z=0$ from the posterior distribution's $1\sigma$ $\gamma$ constraints. We see that broader uncertainty on $\gamma$ in the pseudo model manifests on linear scales, however for $k>1~h/\mathrm{Mpc}$ the variation in the power spectrum becomes larger for $\gamma+q_1$ than for $\gamma$ only. This also implies that for more conservative scale-cuts we should not obtain differences in $\sigma_\gamma$, this effect is purely due to the inclusion of nonlinear scales and high sensitivity to these scales in a Stage-IV-like setup. As an additional proof of this conclusion we demonstrate the $\Omega_{\rm m}-\gamma$ contours for stricter scale-cuts of $\ell_{\rm max}=1000$ in the right panel of Fig.~\ref{fig:LCDMboost}. The $1\sigma$ ranges are visibly in better agreement between the models now. The inclusion of a baryonic feedback model increases the errors for the pseudo model to $\sigma_\gamma = 0.04$ and $\sigma_\gamma \approx 0.05$ for the screened model. In this case the strong degeneracy between the screening scale and baryonic parameters explains a more significant broadening of the contours. We study this degeneracy in detail in Section~\ref{sec:GQbar}.

\subsection{Screening Scale with nDGP Gravity}
\label{sec:DGP_QG}
\begin{figure}
    \centering
    \includegraphics[width=0.49\columnwidth]{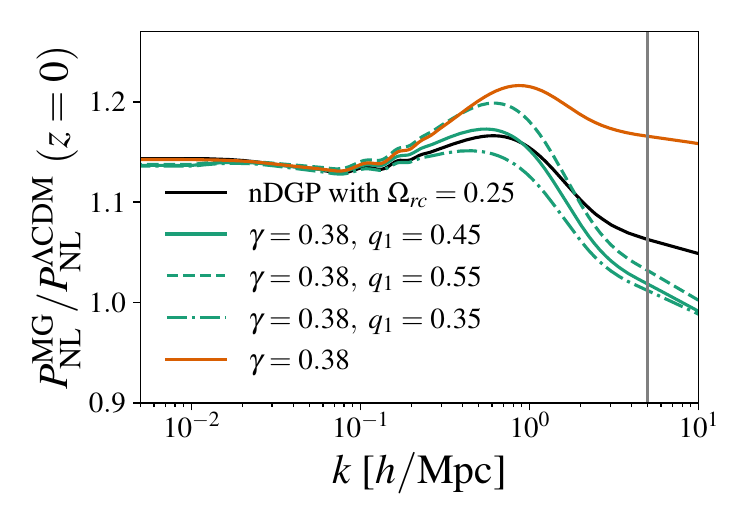}
    \includegraphics[width=0.49\columnwidth]{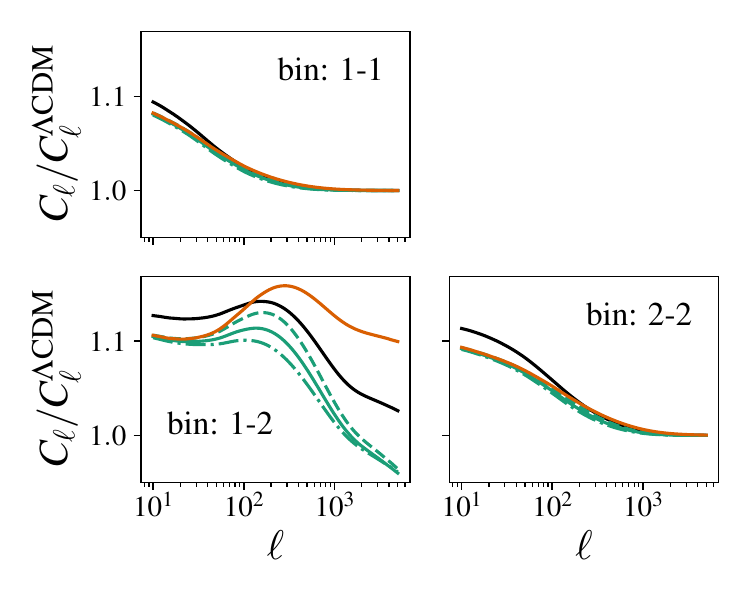}
    \caption{\textit{Left panel}: Power spectrum ratios at redshift $z=0$ for screened (green lines) and unscreened (orange line) models with nonlinear prescriptions described in the text. Grey line denotes $k_{\rm max}=5~h/\mathrm{Mpc}$ that we compute and emulate with \texttt{ReACT}, for higher values we interpolate the MG boost as a power law. \textit{Right panel}: Corresponding ratios of the shear angular power spectra. Central redshifts: for bin 1 $z_c=0.21$, for bin 2 $z_c=0.49$. We fix all cosmological parameters to the fiducial values.}
    \label{fig:screening}
\end{figure}

Next we validate our model on a theory with the Vainshtein-screening mechanism on nonlinear scales and a strong deviation from GR on linear scales ($\sim 14\%$ more structure than in GR on linear scales at lower redshifts). In the left panel of Fig.~\ref{fig:screening} we show the ratio of power spectra $P^{\rm MG}_{\rm NL}/P^{\Lambda \rm CDM}_{\rm NL}$ at $z=0$ for the nDGP model with $\Omega_{rc}=0.25$ (black solid line). For the same fiducial cosmological parameters, we show also this ratio for the screened (green line) and pseudo (orange line) growth index models. The pseudo approach captures the bump-feature in the range of $k \sim 0.2-0.9~h/$Mpc but overpredicts its amplitude. The screened model reproduces this feature extremely well, at the percent level for $k\lesssim 3~h/\rm{Mpc}$. In the right panel, we show ratios of the corresponding angular power spectra, $C_\ell$, for the 3 models with respect to GR for the same set of cosmological parameters. The bump-feature in the nonlinear regime is visible in cross-correlated redshift-bins, while in the auto-correlated bins we see characteristic curves that tightly converge for all models at smaller scales. However, we notice a shift at linear scales for the $\gamma$ models. The reason for this shift is the difference in the time evolution between nDGP and the growth index parametrisation as discussed in Section~\ref{sec:MGscenarios}: while the growth modification is matched between different models at $z=0$, this is no longer the case for higher redshifts. 
\begin{figure}
    \centering
    \includegraphics[width=0.6\columnwidth]{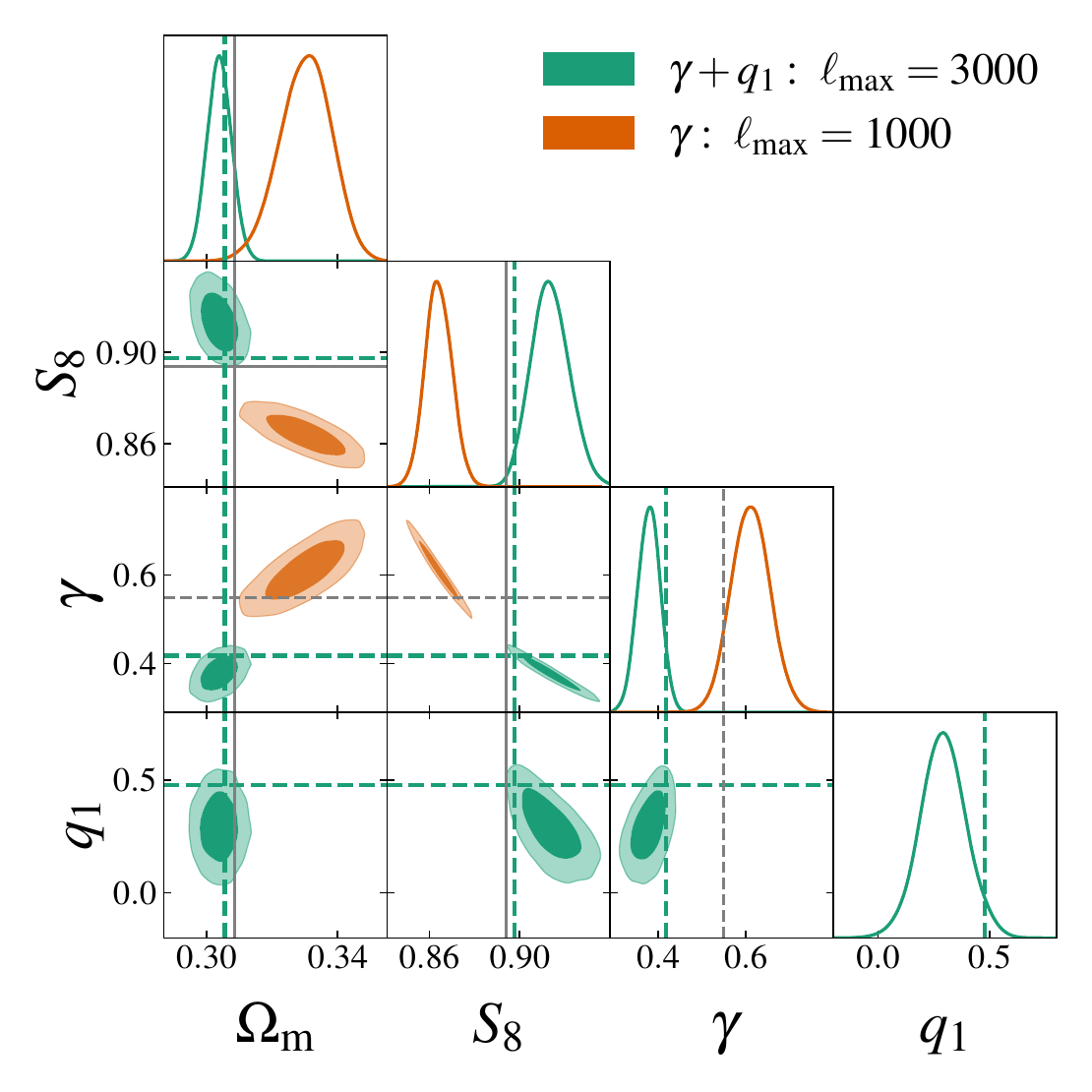}
    \caption{Marginalised posterior distribution for the cosmological
      parameters for the nDGP cosmology and the two model and scale-cut choices,
      as detailed in the legend. The baryonic boost is set to unity. Grey solid lines  mark the
      true values of the synthetic data, grey dashed lines mark the GR-limit. Green dashed lines denote the best-fit values from the MCMC chain for $\gamma+q_1$ model.}
    \label{fig:DGP-GQ-cosmo}
\end{figure}

We consider the same two models: models with screening in the reaction function ($\gamma+q_1$) and without screening via a pseudo-power spectrum ($\gamma$) and repeat our analysis on the mock nDGP data with $\ell_{\rm max}=3000$ and using the same setup. We show the marginalised posterior distribution in Fig.~\ref{fig:DGP-GQ-cosmo}. The full posterior distribution for the sampled cosmological parameters is shown in Fig.~\ref{fig:nDGP_Gmodels_full} of Appendix~\ref{sec:nDGPdiscussion} with a more detailed description. Overall, our model with screening  (green contours and lines) correctly captures the fiducial cosmology, detects a $\gamma$-value lower than its GR-limit as expected, namely $\gamma = 0.38\pm 0.03$, and detects a screening transition with $q_1 = 0.29 \pm 0.10$. We notice a $1\sigma$-bias in $\Omega_{\rm m}$ and a $2\sigma$-bias in $S_8$. Both are consequences of fitting a model with the different time-evolution evolution of growth to data with small error-bars. We check whether this can also be connected to projection effects \cite{Moretti:2023drg}. In other words whether the likelihood maxima (green dashed lines) are at the true values of cosmological parameters and not at the posterior maximum values. From the MCMC chain, we derive the best-fit value $\Omega^{\rm best-fit}_{\rm m}\approx 0.306$, which is closer to the fiducial value of $\Omega^{\rm fid}_{\rm m}\approx 0.309$ than the posterior mean of $\Omega^{\rm mean}_{\rm m} \approx 0.304$. Also the best-fit values for $\gamma^{\rm best-fit}\approx 0.42$ and $S_8^{\rm best-fit} \approx 0.898$ very close to the true value of $S_8^{\rm nDGP} \approx 0.895$. We use the likelihood minimiser \texttt{minuit}\footnote{\href{https://github.com/jpivarski/pyminuit}{github.com/jpivarski/pyminuit}} and obtain similar values for the best-fit parameters. This implies a presence of projection or prior-volume effects. Lastly, we note that this bias vanishes when more parameters are varied and the contours are broadened (see Section~\ref{sec:DGPall}). Overall, the agreement within $2\sigma$ satisfies the  ``precision versus accuracy'' test for our model.

\begin{figure}
    \centering
    \includegraphics[width=0.6\columnwidth]{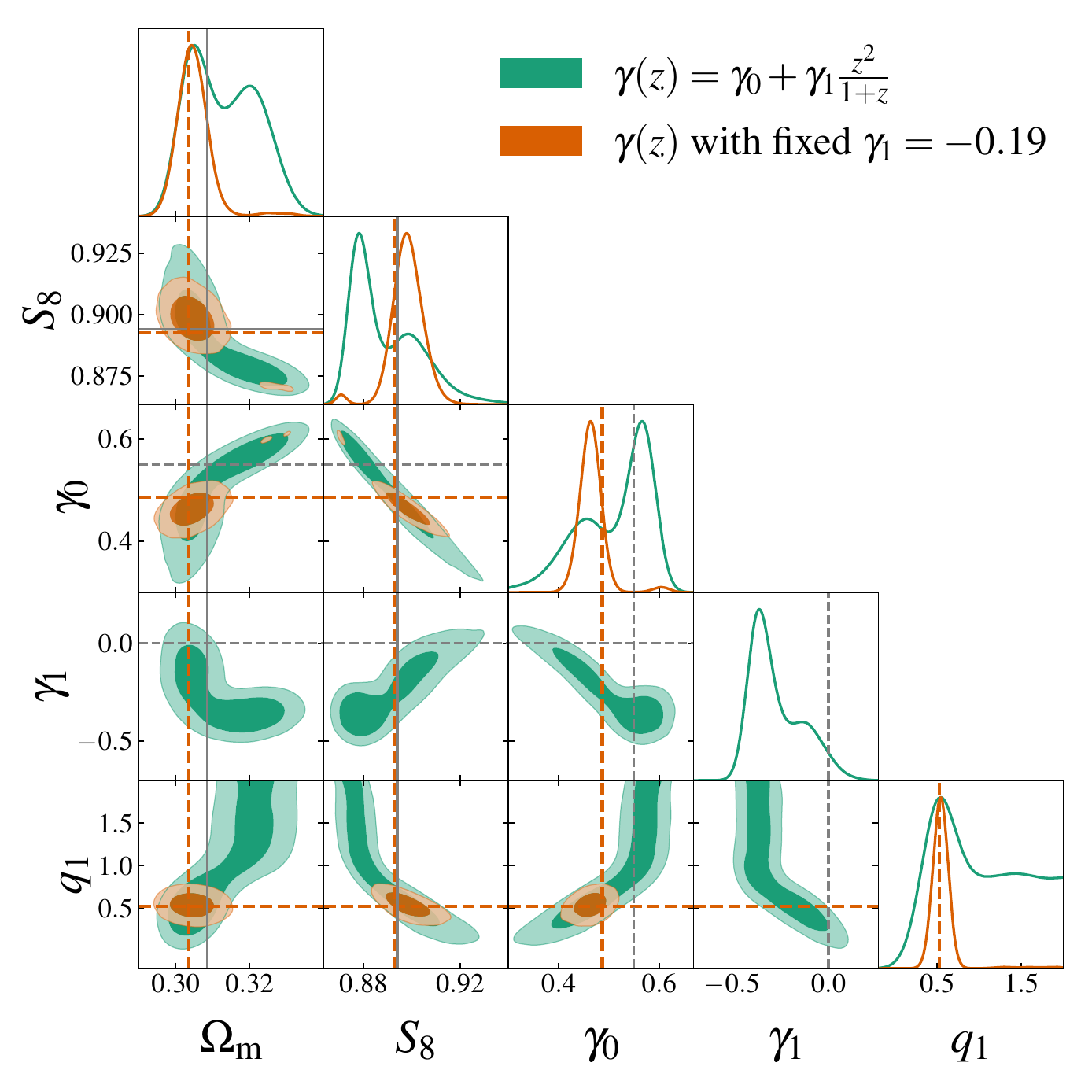}
    \caption{Marginalised posterior distribution for the cosmological and extended
      parameters for the analysis with the fiducial nDGP cosmology using the cosmic shear power spectrum with $\ell_{\rm max}=3000$. We fit the time-dependent growth index with the screening scale, with and without fixing a value of $\gamma_1$, as detailed in the legend. Grey solid lines mark the true values of the synthetic data, grey dashed lines mark the
GR-limit. Orange dashed lines denote the best fit values from the MCMC chain for the second model.}
    \label{fig:DGP-GQz-cosmo}
\end{figure}

In Fig.~\ref{fig:DGP-GQ-cosmo}, we notice that even for scale-cuts below the pessimistic scenario for Stage-IV surveys, namely for $\ell_{\rm max}=1000$, there is a significant bias in $S_8$ and $\Omega_{\rm m}$ for the model without screening (orange contours and lines), with $S_8$ being closer to its GR value. Moreover, the value of $\gamma$ does not correspond to the expected behaviour on large scales but rather points towards suppression of structure formation relative to the standard cosmology. This signals that ignoring a correct screening implementation will result in the wrong extracted cosmology.

In the nonlinear regime our model-independent screening scale seems to be in a good agreement with nDGP theory which includes Vainshtein screening mechanisms. The disadvantage of this model is the inaccurate representation of time evolution of growth on linear scales. We can improve this by introducing the time-dependent gamma-parametrisation from Ref.~\cite{Wen:2023} (also see Section~\ref{sec:MGscenarios}). The results are shown in Fig.~\ref{fig:DGP-GQz-cosmo} and Fig.~\ref{fig:nDGP_GQz_full} for the same setup as before. From the green contours in Fig.~\ref{fig:DGP-GQz-cosmo} we see that while fiducial values lay within $1\sigma$ of all parameters, the screening scale is not properly constrained with cosmic shear information alone. Note how in general, extended parameters make the posterior distribution non-Gaussian by bringing additional degeneracies with cosmological parameters and with each other. Two extended parameters $\gamma_0$ and $\gamma_1$ are strongly anti-correlated, so we can aim to break degeneracies by fixing $\gamma_1$. We find from fitting to the boost at lower redshifts that $\gamma_1=-0.19$ is a good fit and it breaks the degeneracy and allows us to constrain the screening parameter $q_1= 0.57^{+0.07}_{-0.12}$ with $\gamma_0 = 0.47 \pm 0.02$ (orange contours and lines). In Fig.~\ref{fig:DGP-GQz-cosmo}, we clearly see that while $\Omega_{\rm m}$ is still $1\sigma$-biased towards lower values, the bias in $S_8$ towards higher values decreases to $1\sigma$ contrary to the time-independent growth index. We again compute and plot best-fit values (dashed orange lines) from the MCMC chain, for fixed $\gamma_1$ they are $\Omega^{\rm best-fit}_{\rm m}\approx 0.304$ ($0.306$ from \texttt{minuit}), $\gamma^{\rm best-fit}\approx 0.49$, and $S_8^{\rm best-fit} \approx 0.894$. The model with more accurate time-evolution of growth seems to decrease the projection effects. The necessity of tight constraints on $\gamma_1$ implies that additional measurements from clustering will improve our constraints. We aim to investigate probe combinations in future work.

\subsection{Screening Scale versus Baryons}
\label{sec:GQbar}

\begin{figure}
    \centering
    \includegraphics[width=0.49\columnwidth]{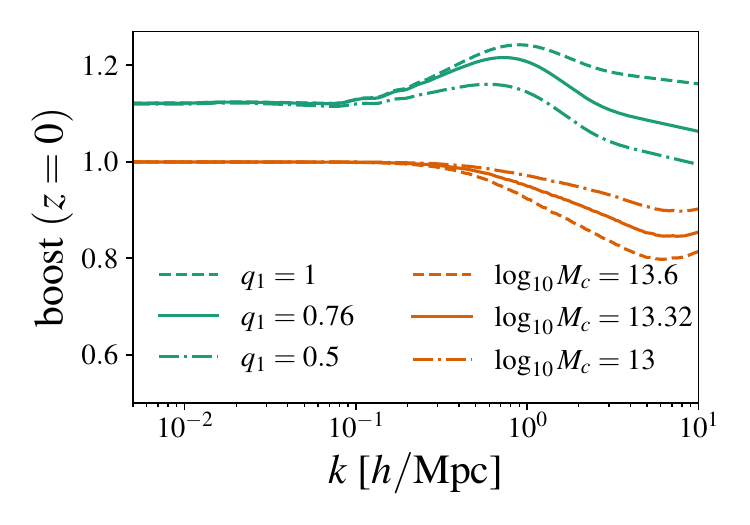}
    \includegraphics[width=0.45\columnwidth]{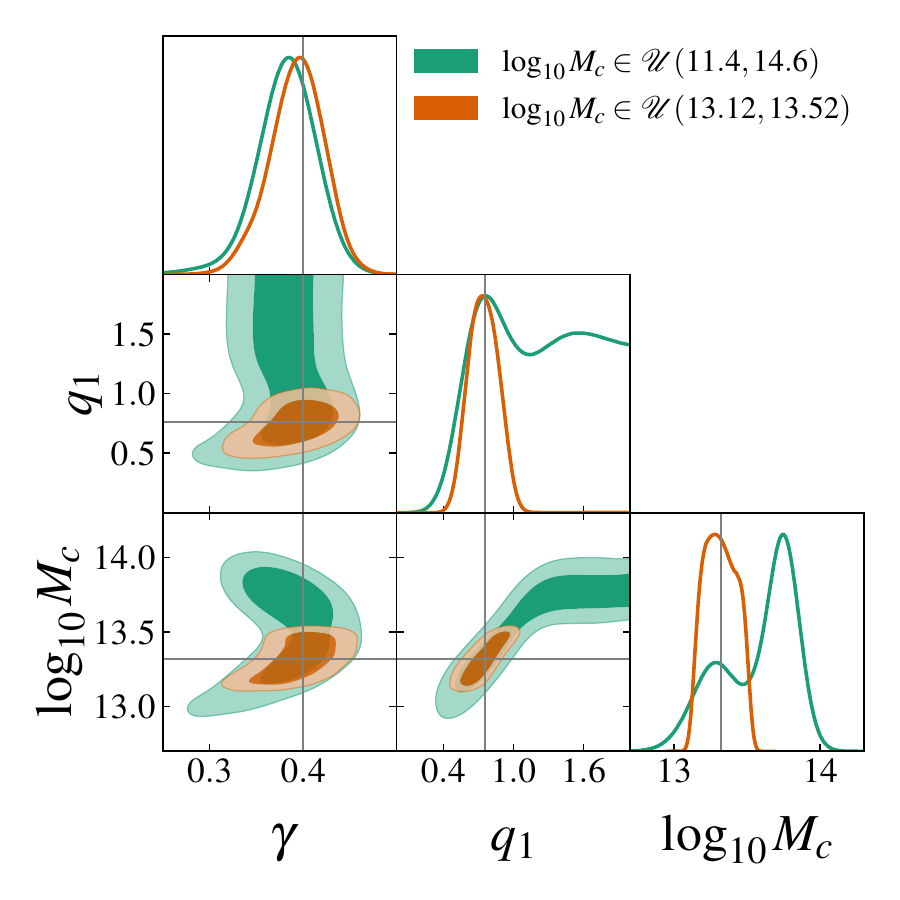}
    \caption{\textit{Left panel}: the power spectrum boost at redshift zero for the MG contribution, $\gamma=0.4$ and varying $q_1$ (green lines), and baryonic feedback (orange lines). \textit{Right panel}: marginalised posterior distributions from the analysis with the fiducial $\gamma+q_1$ cosmology using the cosmic shear power spectrum with $\ell_{\rm max}=3000$ for the MG and baryonic parameters with different priors on the latter, as detailed in the legend. Grey lines mark the true values of the synthetic data.}
    \label{fig:GQ-bar}
\end{figure}

Here we explore the degeneracy between the suppression due to the screening mechanism and baryonic feedback effects. 
In the left panel of Fig.~\ref{fig:GQ-bar} we demonstrate the comparison between MG and baryonic boosts for different parameters. The larger the value of $q_1$, the less is the suppression due to screening. The larger the value of $\log_{10}{M_c}$, the stronger is the suppression due to baryonic feedback at nonlinear scales. It is clear that both effects become noticeable at the same scales $k \gtrsim 1 ~h/\mathrm{Mpc}$, since they demonstrate similar scale-dependence and overall amplitude. Already from this we can conclude that our model-independent screening scale, $q_1$, is highly degenerate with the baryonic suppression parameter $\log_{10}{M_c}$. 

In order to avoid biases due to different time evolution of growth structure on linear scales, we perform an MCMC-analysis on the mock data computed with the $\gamma+q_1$ model directly with $\gamma^{\rm fid}=0.4$ and $q_1^{\rm fid}=0.76$. We show the marginalised posterior distribution for $\gamma-q_1-\log_{10}{M_c}$ in the right panel of Fig.~\ref{fig:GQ-bar}, while the corresponding full posterior distribution is shown in Fig.~\ref{fig:GQ_bar_full}. The unbiased determination of $\gamma$, as well as other cosmological parameters, is not affected by the inclusion of the baryonic feedback. However, the screening scale demonstrates a very strong positive correlation with the baryonic parameter and prefers the unscreened regime of $q_1 >1.5$. The baryonic parameter absorbs the suppression due to the screening and as a result gets biased towards higher values with respect to the fiducial value. This degeneracy persists, when testing variations in the fiducial values of $\log_{10}{M_c}$ within the same $\gamma+q_1$ model, while also examining different fiducial values of $\gamma+q_1$ while keeping $\log_{10}{M_c}$ constant. We perform these checks to mitigate potential coincidences in the choice of fiducial parameters. However, we find that when we impose a tight flat prior on $\log_{10}{M_c}$, the screening scale is recovered unbiased. This means we require a good understanding of baryonic physics for the detection of the model-independent screening transition: for instance, tight priors on baryonic parameters of the order $\sigma_{\log_{10}{M_c}} =0.1$ or $\sigma_{\log_{10}{M_c}}/\log_{10}{M_c} \approx 1\%$.

Instead of imposing priors, for example, based on independent galaxy cluster X-ray observations, a better approach will be the cross correlation or the joint analyses of weak lensing and X-ray/kSZ probes like in the works mentioned in Section~\ref{sec:baryonic_effect}. We also highlight the importance of non-standard hydro-dynamical simulations. The \texttt{BCEMU} emulator we use in this work was trained for standard cosmologies. Despite the flexibility of the baryonification model with 7 or 3 parameters, we still do not know whether this is accurate enough to model nonlinear baryonic physics in beyond-$\Lambda$CDM cosmologies. Reassuringly, Refs.~\cite{Mead:2016a,Arnold:2019zup} show both within the halo model and in $N$-body simulations that baryons and MG physics can be modelled independently to a large extent. A promising argument for this would be that the astrophysics responsible for baryonic effects is happening on the scales in the screened regime, hence there is no reason to believe that it is significantly different from the standard baryon feedback processes. Especially since these are only weakly cosmology dependent (up to several percent for small variations in cosmology) and mainly depend on the fraction of baryons to the total matter according to various studies (e.g., Refs.~\cite{vanDaalen:2011xb, Schneider:2019snl}). However, more work is needed to confirm whether these assumptions satisfy the requirements of Stage-IV surveys \cite{Elbers:2024dad}.

\subsection{Screening Scale versus Neutrinos}
\label{sec:GQnu}

\begin{figure}
    \centering
\includegraphics[width=0.49\columnwidth]{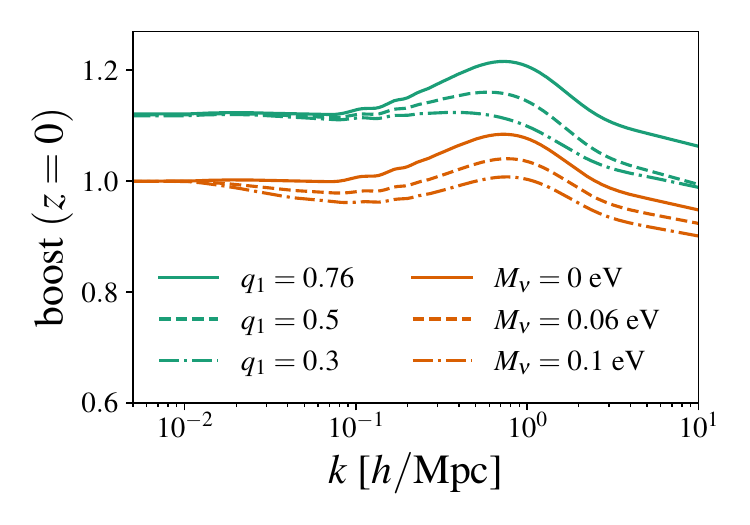}
    \includegraphics[width=0.5\columnwidth]{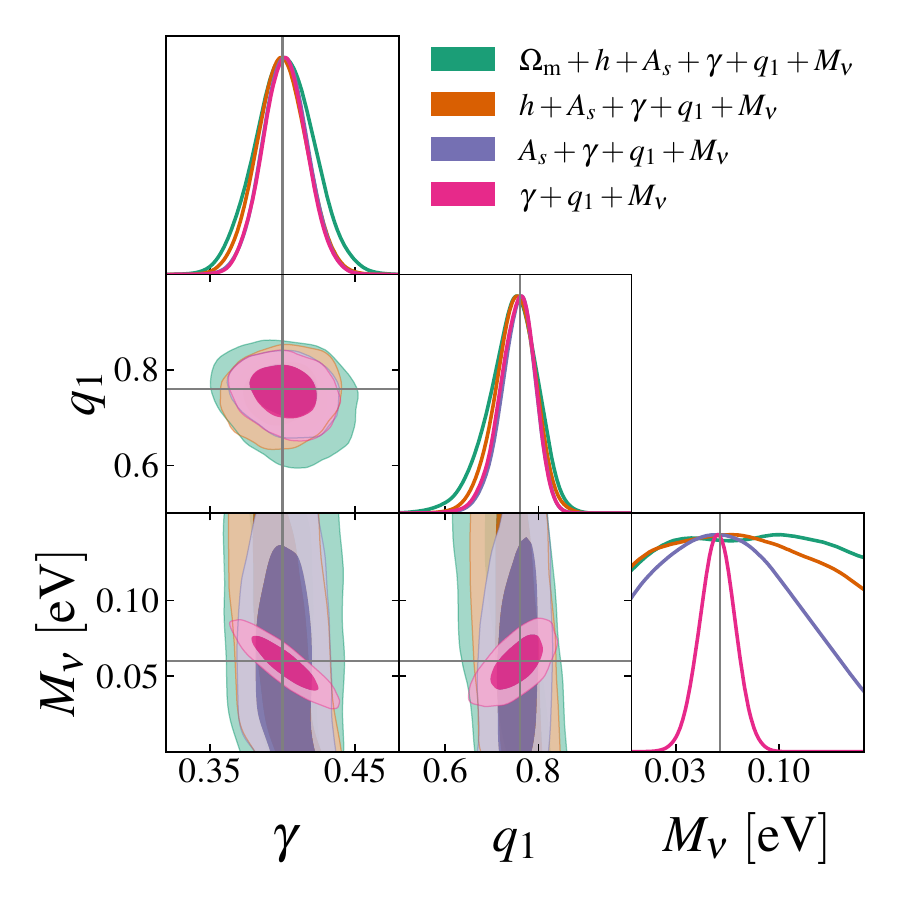}
    \caption{\textit{Left panel}: the power spectrum boost at redshift zero for the MG contribution with $\gamma=0.4$, $M_\nu=0$ and $q_1$ varied (green lines), and $\gamma=0.4$, $q_1=0.76$, $M_\nu$ varied (orange lines). The boosts for various neutrino masses are normalised to the value at the largest scales computed. \textit{Right panel}: marginalised posterior distributions from the analysis with the fiducial $\gamma+q_1$ cosmology using the cosmic shear power spectrum with $\ell_{\rm max}=3000$ for models with different cosmological parameters varied, as detailed in the legend. Grey lines mark the true values of the synthetic data.}
    \label{fig:GQ-nu}
\end{figure}

The last nonlinear effect to consider is the suppression of structure growth due to the contribution of massive neutrinos. In the left panel of Fig.~\ref{fig:GQ-nu} we demonstrate the comparison between MG boosts for different values of screening and neutrino mass parameters. The larger the value of $q_1$, the smaller the suppression due to screening. The larger the value of $M_\nu$, the stronger is the suppression due to massive neutrinos. However, massive neutrinos lead to suppression of growth already at  mildly nonlinear scales $k \gtrsim 0.1 ~h/\mathrm{Mpc}$. Furthermore, the scale-dependency in the suppression due to massive massive neutrinos is drastically different form the suppression due to screening: there is no clear feature in the power spectrum in a constrained scale range. We still expect a positive correlation between $q_1$ and $M_\nu$, but this will be in addition to noticeable degeneracies with other cosmological parameters controlling the amplitude and slope of the power spectrum at all scales, and not only in the nonlinear regime. For instance, the primordial amplitude $A_s$, the expansion rate $h$, and the matter density $\Omega_{\rm m}$. In the right panel of  Fig.~\ref{fig:GQ-nu}, for the same fiducial $\gamma+q_1$ cosmology as in the previous section with the addition of massive neutrinos with $M_\nu^{\rm fid}=0.06$ eV, we showcase how the more parameters we vary the more sensitivity to the neutrino mass we are losing. For example, the prominent and expected negative correlation between $\gamma$ and $M_\nu$ (pink contour) disappears even if just one additional parameter is varied (purple contour). We conclude that Stage-IV cosmic shear measurements alone are not sufficient to put any constraints on the neutrino mass, which is in agreement with the findings of other works, for example Ref.~\cite{SpurioMancini:2023mpt}.

\subsection{Combined Nonlinear Effects}
\label{sec:DGPall}

\begin{figure}
    \centering
    \includegraphics[width=0.8\columnwidth]{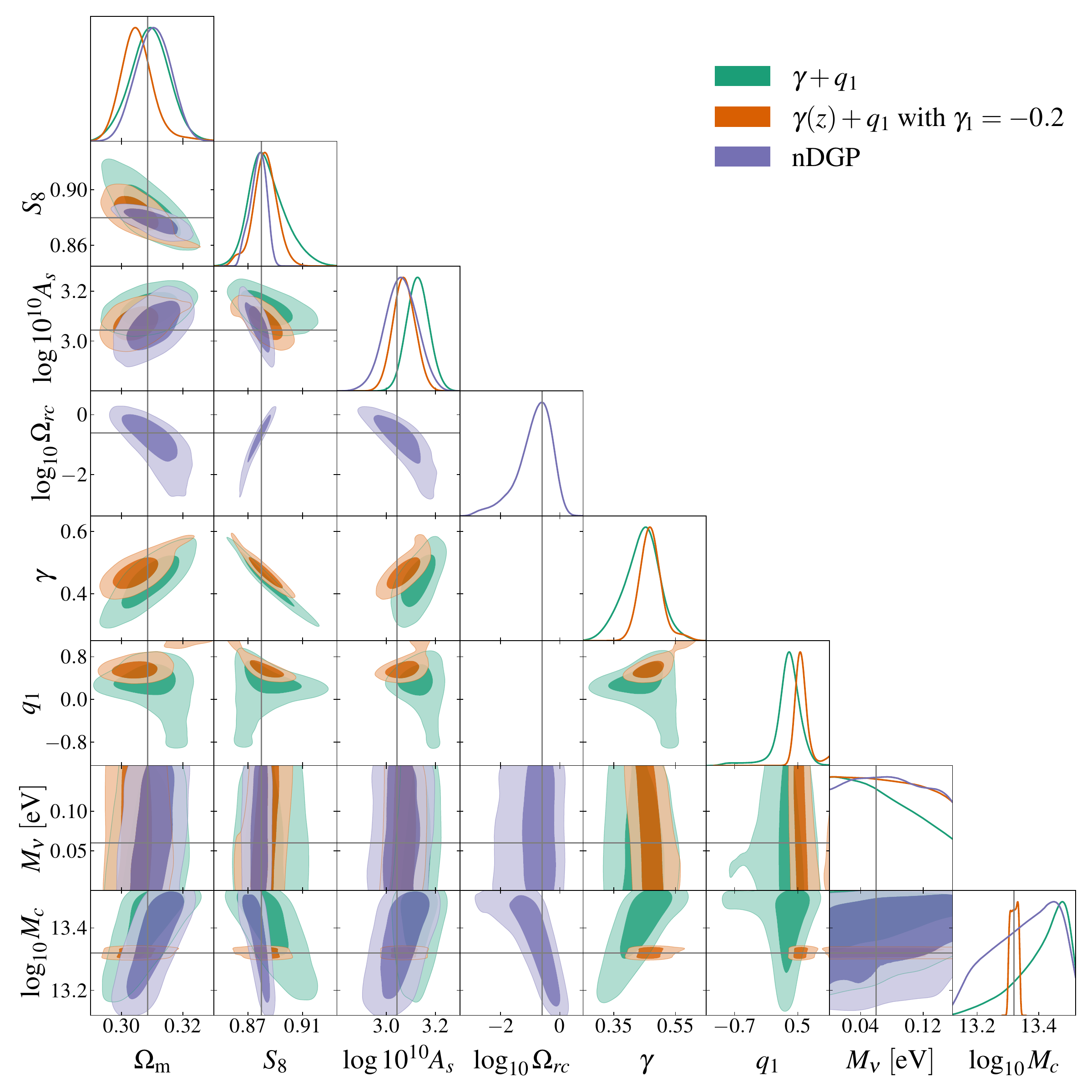}
    \caption{Marginalised posterior distribution for the cosmological and MG parameters for the analysis of the fiducial nDGP cosmology with massive neutrinos and baryonic feedback with a tight uniform prior on the baryonic parameter, using the cosmic shear power spectrum up to $\ell_{\rm max}=3000$. For nDGP and $\gamma+q_1$ the prior on $\log_{10}{M_c}$ is $\mathcal{U}(11.4, 14.6)$, while for the $\gamma(z)+q_1$ model it is $\mathcal{U}(13.3, 13.34)$. Grey lines mark the true values of the synthetic data.}
    \label{fig:DGP-bar-Mnu-GQ-cosmo}
\end{figure}

Finally, we combine our findings in the previous sections to demonstrate how our model-independent approach performs if the underlying cosmology is nDGP with massive neutrinos (with $M_\nu=0.06$ eV) and known baryonic feedback effects (within some prior). All background properties and the growth factor in the IA terms are computed with the total matter density including neutrinos via $\Omega_{\rm m}$, while for the baryonification, since it was trained without neutrinos and is relevant on nonlinear scales, we exclude the neutrinos contribution in $f_b=\Omega_{\rm b}/(\Omega_{\rm m}-\Omega_\nu)$. 

In our main result of this work, Fig.~\ref{fig:DGP-bar-Mnu-GQ-cosmo} (the full posterior distribution can be found in Appendix~\ref{sec:FullPost}), we compare our model independent approach (green colour for time-independent $\gamma$, orange colour for time-dependent $\gamma$) against the exact MG model (purple colour). Overall, for the fiducial value of $\log_{10}{\Omega_{rc}}=-0.6$, we get the mean value of $\log_{10}{\Omega_{rc}}= -0.86^{+0.69}_{-0.36}$, which corresponds to $\gamma= 0.44^{+0.06}_{-0.05}$ with $q_1 =0.33^{+0.21}_{-0.17}$ or $\gamma_0= \gamma = 0.47^{+0.03}_{-0.04}$ with $\gamma_1=-0.2$ and $q_1=0.58^{+0.08}_{-0.14}$. The differences in the $S_8$ uncertainties between models are due to the strong $\gamma$-dependence of the growth at lower redshifts (this is discussed in Appendix~\ref{sec:GRdiscussion}). However, the inferred value of $S_8$ agrees between all models. This shows that the growth index parametrisation seems to be a reasonable approximation in the late universe. It is in the connection to the earlier universe where we see discrepancies in the primordial amplitude, due to different growth evolution. For the time-dependent growth index we infer slightly biased $\Omega_{\rm m}$ but more accurate $\log{10^{10}~A_s}$. The tighter constraint on $\Omega_{\rm m}$ with $\gamma(z)$ is due to tighter prior on the baryonic parameter. In general, we notice that the uncertainties on cosmological parameters are in a good agreement between the exact and model-independent approaches. Therefore, we advocate that our model-independent approach performs as well as the exact modelling, but is more general.

In Fig.~\ref{fig:DGP-bar-Mnu-GQ-cosmo} we also notice that $\log_{10}M_c$ is biased towards higher values in all models. It is highly degenerate with the extended parameters $\log_{10}{\Omega_{rc}}, ~\gamma$, as well as with $\Omega_{\rm m}, ~h, ~\log{10^{10}~A_s}$. We impose an informative prior on the baryon feedback $[\log_{10}{M^{\rm fid}_c}-0.2, \log_{10}{M_c^{\rm fid}}+0.2]$ for nDGP and $\gamma+q_1$ based on the investigation from Section~\ref{sec:baryonic_effect}. For the $\gamma(z)+q_1$ model, we require tighter priors on the baryonic feedback and screening parameters: $[\log_{10}{M^{\rm fid}_c}-0.02, \log_{10}{M_c^{\rm fid}}+0.02]$ and $[-1, 1]$ respectively. Otherwise their prominent degeneracy is weakening the constraints of cosmological parameters, and the screening transition is poorly constrained while preferring larger values closer to the unscreened limit. The total neutrino mass is not constrained in the range of values that we emulated. 

For completeness we list forecasted constraints on the nDGP parameter $\log_{10}{\Omega_{rc}}$ from the literature. In Ref.~\cite{EuclidFrusciante:2023}, for a setup similar to ours and with survey area of $15,000 ~\mathrm{deg}^2$, the authors get from combined cosmic shear, photometric clustering and their cross-correlation analysis with the same fiducial value of $\log_{10}{\Omega_{rc}}$: $\sigma_{\log_{10}{\Omega_{rc}}}= 0.3$ (pessimistic) and $\sigma_{\log_{10}{\Omega_{rc}}}= 0.12$ (optimistic). However, this study does not include baryonic feedback effects, which we demonstrated to be degenerate with the extended parameters. In Ref.~\cite{Harnois-Deraps:2022bie}, for a different setup (5 redshift bins, number of galaxies per arcminute squared per tomographic bin $\bar{n} = 6$, similar shape noise, survey area $5,000 ~\mathrm{deg}^2$) and for a comparably strong modification $\Omega_{rc}=0.36$, the authors find in the optimistic scenario for cosmic shear $\sigma_{\log_{10}{\Omega_{rc}}}= 0.07$, varying three cosmological parameters ($\Omega_{\rm m}, ~S_8, ~h$), and omitting baryonic effects.

\section{Conclusions}
\label{sec:conclusion}

The growth index parametrisation is a single parameter extension of the standard cosmology which allows for deviations in the linear growth functions. Originally developed in the context of spectroscopic measurements, it serves as an indicator of the detection of modified gravity (MG) theories. Previous studies have demonstrated its effectiveness for a few close-positioned redshift bins and up to mildly nonlinear scales \cite{Moretti:2023drg}. In this work, we presented an analysis with this parametrisation extended to the nonlinear regime with a model-independent screening parametrisation, and applied this to cosmic shear forecasts. For weak lensing, most of the information comes from the small, nonlinear scales and is integrated over a broad range of redshifts, in our case $z \in [0.001, 2.5]$.  Our theoretical model for the shear angular power spectrum is based on the halo-based parameterisation introduced in Ref.~\cite{Bose:2022} and was emulated with the {\tt cosmopower} emulator. We also take advantage of the following emulators to accelerate computation of the nonlinear matter power spectrum model: \texttt{BCEMU} for baryonic contribution and \texttt{BACCOemu} \cite{Arico:2021izc} for the linear power spectrum as an input to \texttt{HMcode}.

To validate our model, we first tested it on noiseless mock data generated in the standard ($\Lambda$CDM) cosmology in a Stage-IV setup with a scale cut $\ell_{\rm max}=3000$. We successfully recovered unbiased cosmological parameters and found the expected uncertainty on the growth index, $\gamma$, to be $\sigma_\gamma \approx 0.05$, noting a slight asymmetry in the errors when one baryonic feedback parameter is included. For a model without screening and employing standard structure formation rescaled to match the modifications on large scales, the obtained constraint was $\sigma_\gamma = 0.04$. We further compared the performance and differences between these two $\gamma$-models in the absence of baryonic feedback. We found that the form of our screening model at the nonlinear scales results in tighter constraints on the growth index.

We then conducted a similar analysis on noiseless mock data computed in an nDGP cosmology. This MG theory exhibits enhanced structure growth on linear scales that transitions to GR behaviour on nonlinear scales via a screening mechanism. The analysis was performed on mock data with a rather strong deviation from GR (by $14\%$ at low redshifts in the linear regime). We aimed to demonstrate the robustness of our modelling in this relatively extreme scenario to guarantee its validity in more realistic scenarios with a weaker deviation form the standard cosmology.
We found that our model with screening $\gamma+q_1$ performs well: it successfully recovers cosmological parameters within $2\sigma$, finds $\gamma \sim 0.38$ lower than its GR-limit $\gamma \sim 0.55$, and detects a screening transition with its associated parameter $q_1 \sim 0.29$. 

We also obtained a bias towards higher values in the primordial amplitude $A_{\rm s}$. This bias arises from differences in the time evolution of the linear growth factors between the growth index parametrisation and the nDGP model. This inaccurate representation of the time evolution of structure growth for scalar-tensor theories via a constant growth index parametrisation has been pointed out in Ref.~\cite{Wen:2023}. To address this issue, we explored a time-dependent functional form $\gamma(z)=\gamma_0+\gamma_1 z^2/(1+z)$. Incorporating this time-dependent growth index, we found that the bias in the amplitude vanishes.  However, the constraints on the expansion rate and screening transition are weakened significantly when we considered this model. By fixing $\gamma_1 = -0.19$, we broke the corresponding degeneracies, and we recovered unbiased cosmological parameters with $\gamma_0 \sim 0.47$ and $q_1 \sim 0.57$.

We found that ignoring the screening transition leads to biases in the expansion rate and matter density, as well as to a false detection of $\gamma$ that exceeds its GR-limit. This happens even with scale cuts lower than the ``pessimistic'' scenario in Ref.~\cite{Euclid:2019}, with $\ell_{\rm max}=1500$. This demonstrates the importance of correct inclusion of a screening scale when extended cosmologies are considered.

We proceeded to study the degeneracies between the screening transition parameter $q_1$, baryonic feedback, and massive neutrinos. All three effects are nonlinear and result in the suppression of structure formation at small scales. As before, we explored the case with only one free baryonic feedback parameter $M_c$, which controls the slope of the gas distribution. We found a strong positive correlation between the screening and baryonic parameters. The suppression due to the screening was absorbed by the baryonic feedback if no priors on the latter were imposed. Both effects are prominent in the same range of scales and are highly degenerate. For this reason, we conclude that using cosmic shear alone a detection of the model-independent screening transition is possible only if tight priors on the baryonic parameters are imposed. We used the uniform prior $[\log_{10}{M^{\rm fid}_c}-0.2, \log_{10}{M_c^{\rm fid}}+0.2]$ motivated by cluster measurements in Ref.~\cite{Grandis:2023qwx}. Alternatively, a cross-correlation of cosmic shear with X-ray/kSZ observations can break this degeneracy too. With the total neutrino mass as a free parameter, we found that it cannot be constrained with the cosmic shear data alone. This is in agreement with findings in Stage-III surveys \cite{Hildebrandt:2020} as well as in forecasts for Stage-IV experiments \cite{SpurioMancini:2023mpt}. We also found no strong degeneracy between the screening transition and neutrino mass -- while they have similar impact on structure formation, their scale-dependence and strength of impact differ.

Combining all the aforementioned nonlinear effects, we found that our model-independent approach performs well when compared against the exact modelling, and derived the following constraints in the full analysis on nDGP data with massive neutrinos and baryonic feedback with a narrow flat prior:  $\sigma_{\log_{10}{\Omega_{rc}}} \approx 0.53 ~(88\%)$, $\sigma_\gamma \approx 0.06 ~(13\%)$, $\sigma_{q_1} \approx 0.19  ~(58\%)$.

To conclude, we outline the next necessary steps in preparation for a fully model-independent analysis for beyond-$\Lambda$CDM cosmologies with Stage-IV cosmic shear surveys. 
In order to reduce the error bars on extended and baryonic parameters, we advocate for the combination of cosmic shear with photometric galaxy clustering and the corresponding cross-correlation, i.e. the 3$\times$2-point analysis. Similarly, a combination with spectroscopic galaxy clustering can alleviate the degeneracies
and yield tighter constraints \cite{Euclid:2019}. After demonstrating the robustness of the $\gamma(z)$ approach, we aim to explore a more agnostic approach, that would not require any assumption of the time-evolution in the linear growth. For example, we can bin $\mu(z)$ from Eq.~\ref{eq:mu_L} in redshift directly \cite{Pogosian:2021mcs,Srinivasan:2021gib}. The goal would be to find an optimal binning scheme. Additionally, as demonstrated in Ref.~\cite{Bose:2022}, we can extend our modelling to include not only MG theories but also exotic dark energy models.

\acknowledgments

MT would like to thank Alessandra Silvestri for useful comments and suggestions, as well as for the warm hospitality in Leiden University, where a large part of this project was accomplished. MT's research is supported by a doctoral studentship in the School of Physics and Astronomy, University of Edinburgh. BB was supported by a UK Research and Innovation Stephen Hawking Fellowship (EP/W005654/2). PC’s research is supported by grant RF/ERE/221061. AP is a UK Research and Innovation Future Leaders Fellow [grant MR/X005399/1]. For the purpose of open access, the author has applied a Creative Commons Attribution (CC BY) licence to any Author Accepted Manuscript version arising from this submission.

\bibliographystyle{JHEP}
\bibliography{mybib.bib}

\appendix

\section{Connection between Linder gamma and nDGP}
\label{sec:AppB}

For DGP models without an additional dark energy component, we have $\gamma_{\rm DGP} \approx 11/16$  (e.g., Ref.~\cite{Linder:2007}). Here we will derive the value for when we include a dark energy contribution. Following the same steps as in Ref.~\cite{Linder:2007}, one can arrive to the following expression for the growth index by solving the linearised growth equation, taking a matter dominated era limit of $\Omega_{w}(a)/\Omega_{\rm m}(a) \ll 1$, $\Omega_w$ being the dark energy density fraction, and $(\mu_{\rm L}-1) \ll 1$ (Eq. 22 in Ref.~\cite{Linder:2007}):
\begin{align}
    \gamma = \frac{1}{2}+\frac{1}{4\Omega_{w}(a)}\int_0^1\frac{\mathrm{d}u}{u}u^{5/2}\Omega_{w}(au) -\frac{3}{2 \Omega_{w}(a)}\int_0^1\frac{\mathrm{d}u}{u}[\mu_{\rm L}(au)-1]u^{5/2} \, , 
    \label{eq:gammalinearised}
\end{align}
with $\Omega_{w}(a)=1-\Omega_{\rm m}(a)$. For early times $ \Omega_{w}(a)\propto a^{-3w}$ and $\mu_{\rm L}-1=A \, \Omega_{w}(a)$, with $A$ being a parameter to be determined. We can integrate the equation for $\gamma$ to get 
\begin{equation}
\label{eq:gamma_approx}
\gamma = \frac{3(1-w-A)}{5-6w} \, , 
\end{equation}
where for GR $w=-1$ and $A=0$, so one obtains the standard result of $\gamma_{\rm GR}=6/11$.

Now, the Friedman equation for flat DGP models is given by (``+'' for the normal branch or nDGP, ``-'' for the self-accelerating branch or sDGP):
\begin{equation}
\label{eq:eq1}
     H^2 \pm \frac{H}{r_c}=\frac{8\pi G}{3} \rho \, .
\end{equation}
We have 2 options: assume that a) $\rho=\rho_{\rm m}$, hence the additional $H/r_c$  acts like an effective dark energy component $\rho_w=\mp \frac{3}{8\pi G} \frac{H}{r_c}$; or b) $\rho=\rho_{\rm m}+\rho_{\rm DE}$, hence the effective dark energy has two contributions $\rho_w = \rho_{\rm DE} \mp \frac{3}{8\pi G}\frac{H}{r_c}$. 

For the first case without dark energy we find from the Friedman equation: $\Omega_{\rm m}(a)=1 \pm 2\sqrt{\Omega_{rc}}/E(a)$
 or $E(a)=\mp \sqrt{\Omega_{rc}}+\sqrt{\Omega_{rc}+\Omega_{\rm m}a^{-3}}$. The former implies that from $E(a=1)=1$ follows  $\mp 2\sqrt{\Omega_{rc}}=1-\Omega_{\rm m}$. Hence, for all realistic $\Omega_{\rm m}<1$ only the self-accelerating branch is relevant.
From $\dot{\rho}_w=-3H(1+w) \rho_w$ and a derivative of Eq.~\ref{eq:eq1} we find $w= -1/(1+\Omega_{\rm m}(a))$. From Eq.~\ref{eq:beta} and \ref{eq:mu_L} we compute $\beta=-\frac{1+\Omega^2_{\rm m}(a)}{1-\Omega^2_{\rm m}(a)}$ and $\mu_{\rm L}-1 = - \frac{1}{3}\frac{1-\Omega^2_{\rm m}(a)}{1+\Omega^2_{\rm m}(a)}=A\Omega_w(a)$ with $A=- \frac{1}{3}\frac{1+\Omega_{\rm m}(a)}{1+\Omega^2_{\rm m}(a)}$. Combining these findings together with the limit of $a\rightarrow 0$, i.e. $A \rightarrow -1/3$ and $w \rightarrow -1/2$, one gets $\gamma=11/16$ from Eq.~\ref{eq:gamma_approx} exactly like in Ref.~\cite{Linder:2007} (recall Eq.~\ref{eq:gammalinearised} is derived for early times).

Overall, for small values of $\Omega_{rc}$, $\gamma=11/16$ is a good approximation at all redshifts for this particular case of the flat DGP or sDGP model, where the value of the extended parameter is directly related to the matter density. While for larger values of $\Omega_{rc}$ (hence smaller values of $\Omega_{\rm m}$) as $a\rightarrow 1$, $A \rightarrow -1/3$ and $w \rightarrow -1$ (mimicking a cosmological constant today). This decreases the value of the growth index to $\gamma \rightarrow 7/11$, when computed according to Eq.~\ref{eq:gamma_approx}. Even the most extreme values of $\Omega_{rc}$ show at most $2\%$ deviation at lower redshifts when compared against solutions from numerically equating $\mu^\gamma_{\rm L}=\mu^{\rm DGP}_{\rm L}$ at each redshift individually. However, note that in this scenario the growth index value is always larger than its GR limit. In other words, the structure growth is suppressed relative to the standard cosmology.

For the second case with the cosmological constant, $\Omega_{rc}$ becomes an independent parameter. We discuss only the case of only including a cosmological constant, but the same logic can be applied to  any other parametrisation of the dark energy component. We fix the expansion to the $\Lambda$CDM cosmology: from Eq.~\ref{eq:eq1}, we have $E(a)=\mp \sqrt{\Omega_{rc}}+\sqrt{\Omega_{rc}+\Omega_{\rm m}a^{-3} +\Omega_{\rm DE}(a)}$, and from the equality $E=E_{\Lambda \rm CDM}$ we get $\Omega_{\rm DE}(a)=(1-\Omega_{\rm m})\pm 2\sqrt{\Omega_{rc}}\sqrt{\Omega_{\rm m}a^{-3}+(1-\Omega_{\rm m})}$. From Eq.~\ref{eq:beta} and \ref{eq:mu_L}: $\mu_{\rm L}-1 = \frac{2 \sqrt{\Omega_{rc}}}{3} \frac{1}{2\sqrt{\Omega_{rc}} \pm E(a)[2-\Omega_{\rm m}(a)]}$ with ``+'' for nDGP and ``-'' for sDGP. The corresponding limits are: $\mu_{\rm L}-1 \rightarrow \pm \frac{2\sqrt{\Omega_{rc}}}{3\sqrt{\Omega_{\rm m}}}a^{3/2}$ for early times, while the late-time limit tends to $\mu_{\rm L}-1 \rightarrow \frac{2\sqrt{\Omega_{rc}}}{3}\frac{1}{2\sqrt{\Omega_{rc}} \pm [2-\Omega_{\rm m}]}$. This means that at high redshift the assumption of $(\mu_{\rm L}-1)/\Omega_w(a) \approx \mathrm{const}$ is not valid in this case as $\Omega_w(a) \propto a^3$, while $(\mu_{\rm L}-1) \propto a^{3/2}$. As a consequence the 
integration in Eq.~\ref{eq:gamma_approx} for the early-time limit yields a diverging value of $\gamma = \gamma_{\rm GR} \mp \frac{\sqrt{\Omega_{rc}\Omega_{\rm m}}}{4(1-\Omega_{\rm m})a^{3/2}}$, with ``-'' for nDGP, ``+'' for sDGP. We can repeat this calculation for $a\rightarrow 1$ and get $\gamma = \gamma_{\rm GR} - \frac{2}{5}\frac{\sqrt{\Omega_{rc}}}{(2\sqrt{\Omega_{rc}}\pm[2-\Omega_{\rm m}])(1-\Omega_{\rm m})}$ , with ``+'' for nDGP, ``-'' for sDGP. Therefore, the addition of the dark energy component in the nDGP model leads to $\gamma \leq 0.55$ for any $\Omega_{rc} \geq 0$, resulting in enhanced growth at linear scales and in agreement with our findings. We can insert our values for $\Omega_{rc}$ and $\Omega_{\rm m}$ for the late-time limit and obtain $\gamma \approx \gamma_{\rm GR}-0.11 \approx 0.44$, which is in excellent agreement with our findings in Sections~\ref{sec:DGP_QG} and~\ref{sec:DGPall}.

\section{Validation with GR: full posteriors and discussion}
\label{sec:GRdiscussion}

\begin{figure}
    \centering
    \includegraphics[width=\columnwidth]{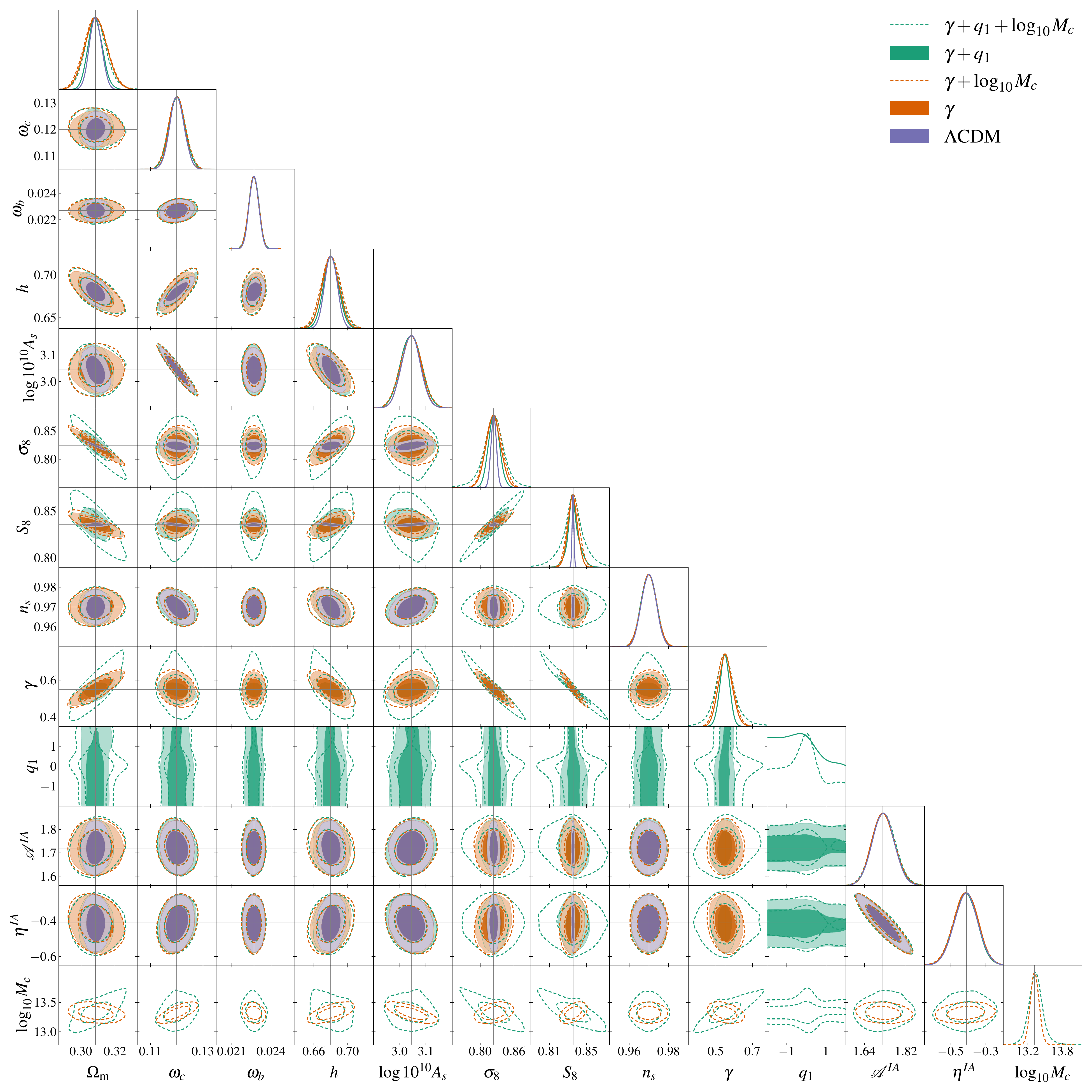}
    \caption{Full posterior distribution for validation with GR mock data for a Stage-IV cosmic shear setup with $\ell_{\rm max}=3000$. Different colours correspond to the growth index model with screening (green), the growth index model with the unmodified nonlinear growth (orange), and the model used to produce the mock data (purple). The dashed lines denote models with an additional baryonic feedback parameter.}
    \label{fig:LCDM_valid_full}
\end{figure}

In Fig.~\ref{fig:LCDM_valid_full} several degeneracies are apparent due to the fact that $C_\ell \propto \sigma_8^2 \Omega_{\rm m}$: the negative correlations of $\sigma_8-\Omega_{\rm m}$ and $\gamma-\sigma_8$, and the positive correlation of $\gamma-\Omega_{\rm m}$. We overlay contours from the $\Lambda$CDM modelling to showcase the strong positive correlation between $\log{10^{10}~A_s}$ and $\sigma_8$, which is weakened when an additional parameter controlling the amplitude of the power spectrum on large scales, $\gamma$, is added. However, in all scenarios $\log{10^{10}~A_s}$ is strongly anti-correlated with $h$ and $\omega_c$. We also note that the zNLA parameters $\mathcal{A}^{\rm IA}$ and $\eta^{\rm IA}$ are not strongly degenerate with any other parameters but are anti-correlated with one another; their constraints are model-independent. From the full posterior we also see that the screening scale is not detected, which is to be expect for $\mu_L^{\gamma \sim 0.55}\approx1$ (see Eq.~\ref{eq:muNL}). 
Additionally, the growth index is anti-correlated with the expansion rate $h$, which is due to the inclusion of large scales, $10<\ell<100$, and tight informative priors on $\omega_b$ and $n_s$ (see discussion in Appendix~\ref{sec:AppDeg}).
\begin{figure}
    \centering
    \includegraphics[width=\columnwidth]{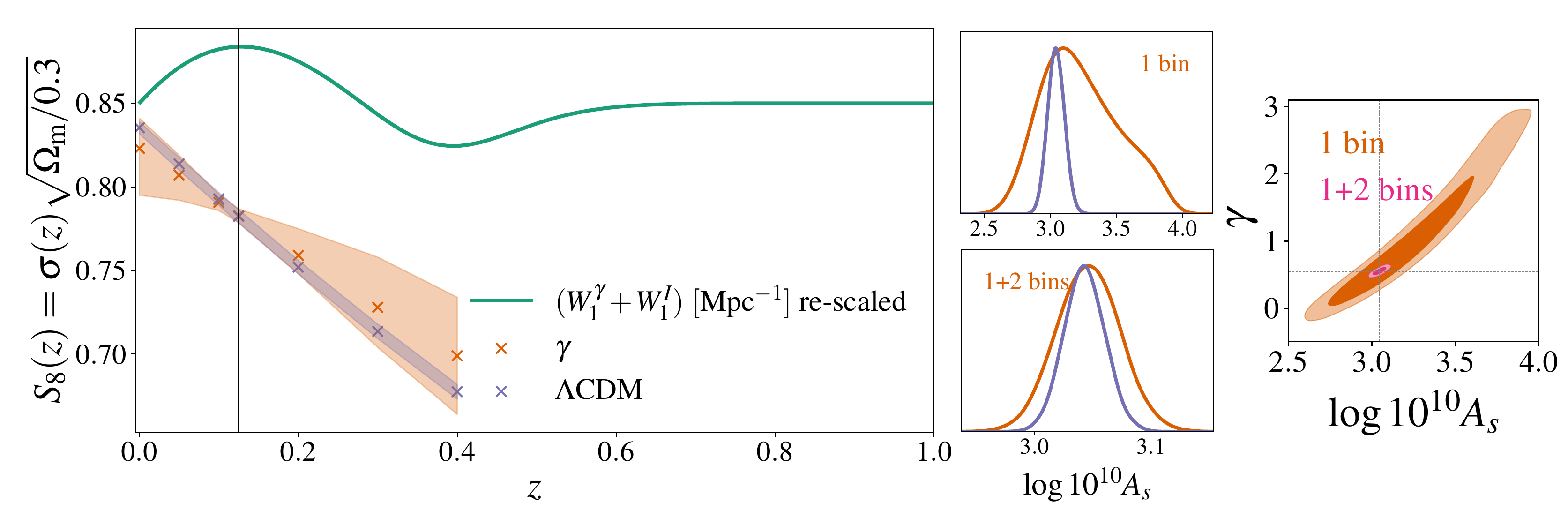}
    \caption{Varying only $\Omega_{\rm m}, ~S_8, ~\gamma$ with re-scaled covariance and using linear scales. \textit{Left panel}: Constraints from the first redshift bin. Time evolution of $S_8(z)$ in the standard cosmology (purple crosses) and $\gamma$-parametrisation ($\gamma$-pseudo, orange crosses). Green solid line denotes the combination of the lensing and IA kernels (re-scaled and shifted for visualisation purposes). Black solid line denotes the maximum of the kernel and same constraints on $S_8(z)$ in both models. \textit{Middle panel}: marginalised posterior distributions for $\log{10^{10}~A_s}$ from the first bin (upper plot) and combined first two bins (lower plot) in the standard cosmology (purple lines) and $\gamma$-parametrisation (orange lines). \textit{Right panel}: Marginalised posterior distribution for $\gamma$ and $\log{10^{10}~A_s}$ with the first photometric bin (orange contour) and combined with the second photometric bin (pink contour). Dashed lines denote the fiducial values.}
    \label{fig:S8}
\end{figure}

From the same full posterior distribution we also notice that the $\gamma+q_1$ (solid green lines) and $\Lambda$CDM (solid purple lines) models constrain the background cosmological parameters, $\Omega_{\rm m}, ~h$, to the same level of uncertainty. This is well understood since $\gamma$ only impacts the amplitude of the power spectrum and not the background. However, we notice drastically different constraints on  $\sigma_8$ and $S_8$: the contours are much broader in the $\gamma$ models than in the $\Lambda$CDM case. This difference arises due to the strong sensitivity of the linear growth factor to the deviation of the growth index from its GR value at lower redshifts (see Fig.~\ref{fig:MGrowths}). To demonstrate this we fix all parameters in the model to their fiducial values and vary $\Omega_{\rm m}$, $S_8$ and $\gamma$ in the first redshift bin with a re-scaled covariance by a factor of 250 (corresponding to smaller error bars), and using linear scales only, $\ell_{\rm max}=500$. In the left panel of Fig.~\ref{fig:S8} we show that, as expected, both models measure the same value of $S_8(z)=\sigma_8(z)\sqrt{\Omega_{\rm m}/0.3}$ around the peak of the first lensing kernel at $z=0.125$. Therefore, it is the variation in $\gamma$-values that affects the inferred constraints of $S_8$ at redshift $z=0$. For the first redshift bin, the constraints of the primordial amplitude demonstrate the same trend (see the upper middle panel of Fig.~\ref{fig:S8}). This changes as soon as we add additional redshift bins (in the lower middle panel). In Section~\ref{sec:DGP_QG} we discuss that the cross-correlated bins provide the most information on the nonlinear scales as well. Overall, in Fig.~\ref{fig:LCDM_valid_full} with 10 redshift bins the constraints on the primordial amplitude are identical between both models since we match the growth at high redshift. Contrary to Stage-III surveys (see, for example, Ref.~\cite{Joachimi:2020abi}), a Stage IV-like setup constrains the primordial amplitude well due to its wide redshift range and large number of redshift bins.

\section{Expansion rate: note on degeneracies and priors}
\label{sec:AppDeg}
\begin{figure}
    \centering
    \includegraphics[width=0.56\columnwidth]{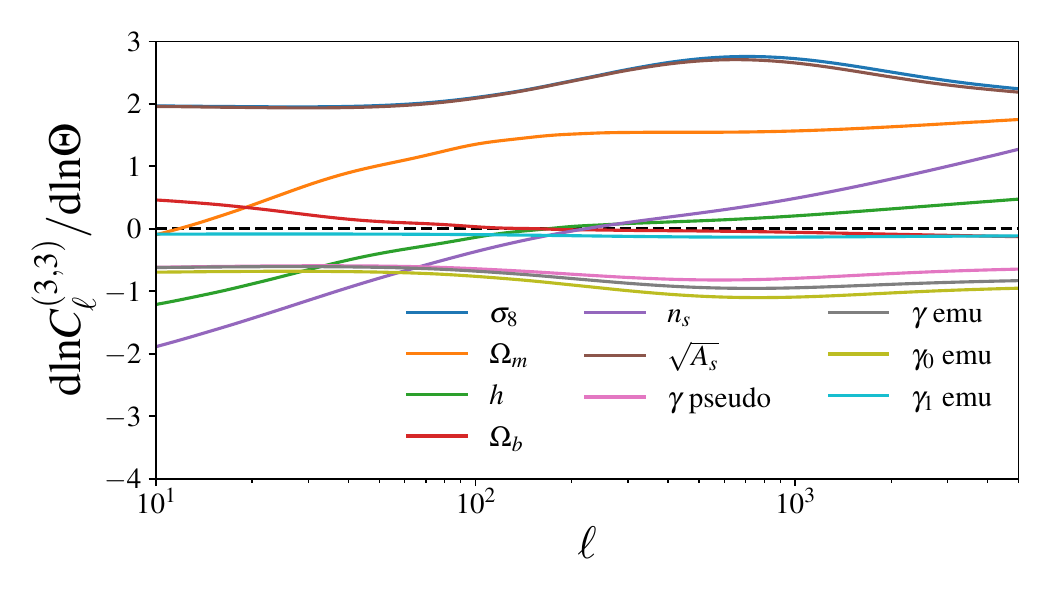}
    \includegraphics[width=0.43\columnwidth]{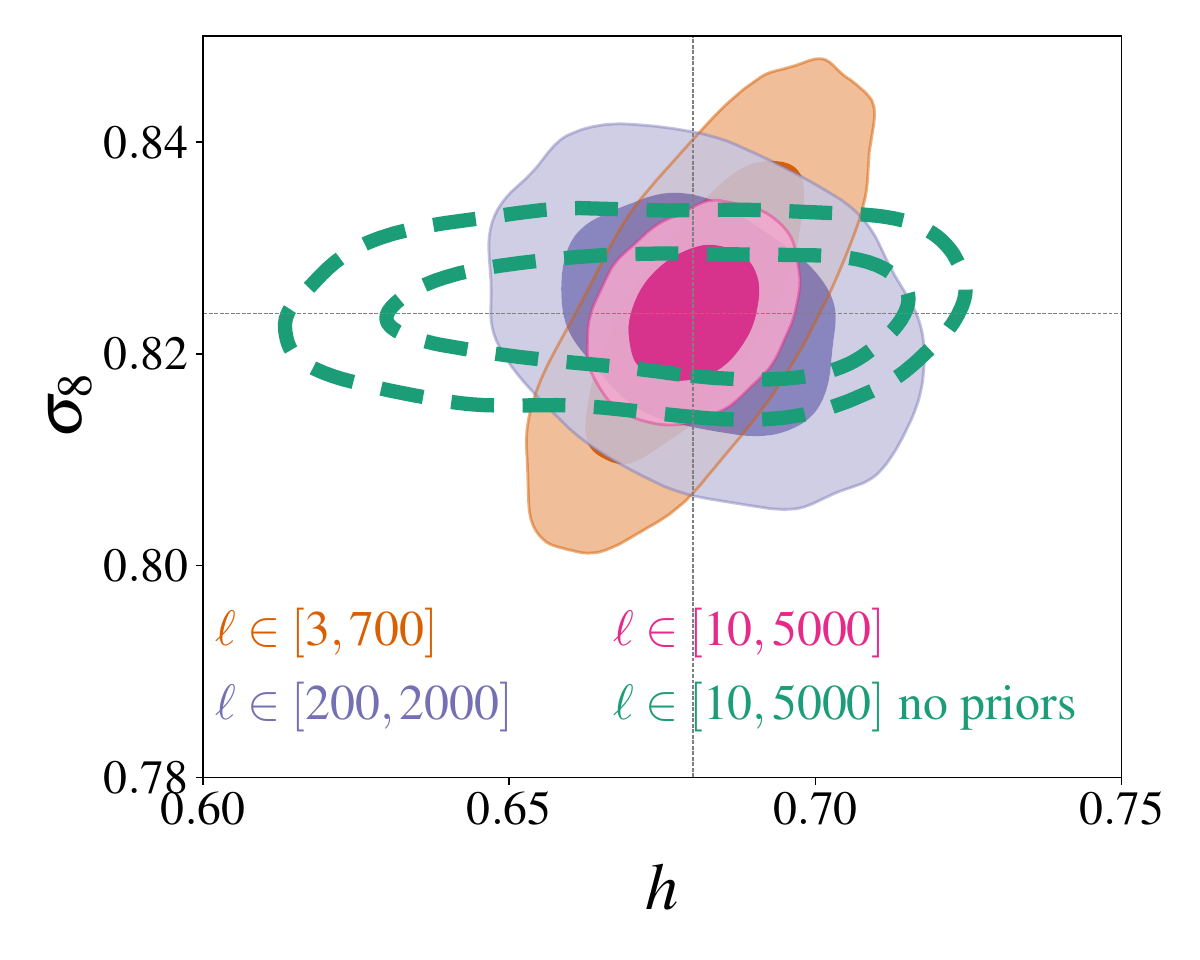}
    \caption{\textit{Left panel}: the parameter dependence of the shear angular power spectrum in the 3-3 redshift bin without shape noise. \textit{Right panel}: change in the orientation of degeneracies between parameters depending on the scale-cuts and priors used.}
    \label{fig:Cell_deg}
\end{figure}
In Fig.~\ref{fig:LCDM_valid_full} we see strong degeneracies between all cosmological parameters and the expansion rate $h$. Fig.~\ref{fig:Cell_deg} shows the derivatives of $C_{\ell}$ in the $3-3$ photo-$z$ bin with respect to the parameters listed in the legend. From  this figure, it is clear that $C_\ell$ is insensitive to $h$ in the region of $10^2<\ell<10^3$. We also note the change in the dependence of $C_\ell$ on $h$ in the  $\ell<10^2$(prominent negative) and $\ell>10^3$ (weak positive) regimes. From Eq.~\ref{eq:Cell_WL}, we see that the impact of $h$ on $C_\ell$ is coming purely from the matter power spectrum. In the linear regime, i.e., low $\ell$ and $k$, the matter power spectrum is an approximate power law with its slope depending on $h$, $n_s$ and $\omega_b$ (for more detailed discussion see Ref.~\cite{Hall:2021qjk}). We impose a Planck prior on $n_s$ and a BBN prior on $\omega_b$, which breaks this degeneracy. In the right panel of Fig.~\ref{fig:Cell_deg}, we demonstrate the rotation of degeneracy between $\sigma_8$ and $h$, when these informative priors are imposed in the analysis with a GR model on GR mock data. For $\ell_{\rm max}=700$, $\mathrm{dln}C_\ell/\mathrm{dln}h$ and $\mathrm{dln}C_\ell/\mathrm{dln}\sigma_8$ have opposite signs, hence the orange contour demonstrates a positive correlation. The situation changes to the negative correlation for $\ell \in [200, 2000]$ (the purple contour). When all scales are combined (the pink contour), the orientation of the degeneracy still slightly prefers the positive correlation characteristic for lower $\ell$-values. However, this is no longer the case, when the priors are omitted (the dashed green contour). Similar arguments are applicable to all other cosmological parameters and their degeneracies with $h$. Also note that due to the choice of a diagonal Gaussian covariance, our constraints can be considered optimistic.

\section{Test with nDGP: full posteriors and discussion}
\label{sec:nDGPdiscussion}

\begin{figure}
    \centering
    \includegraphics[width=\columnwidth]{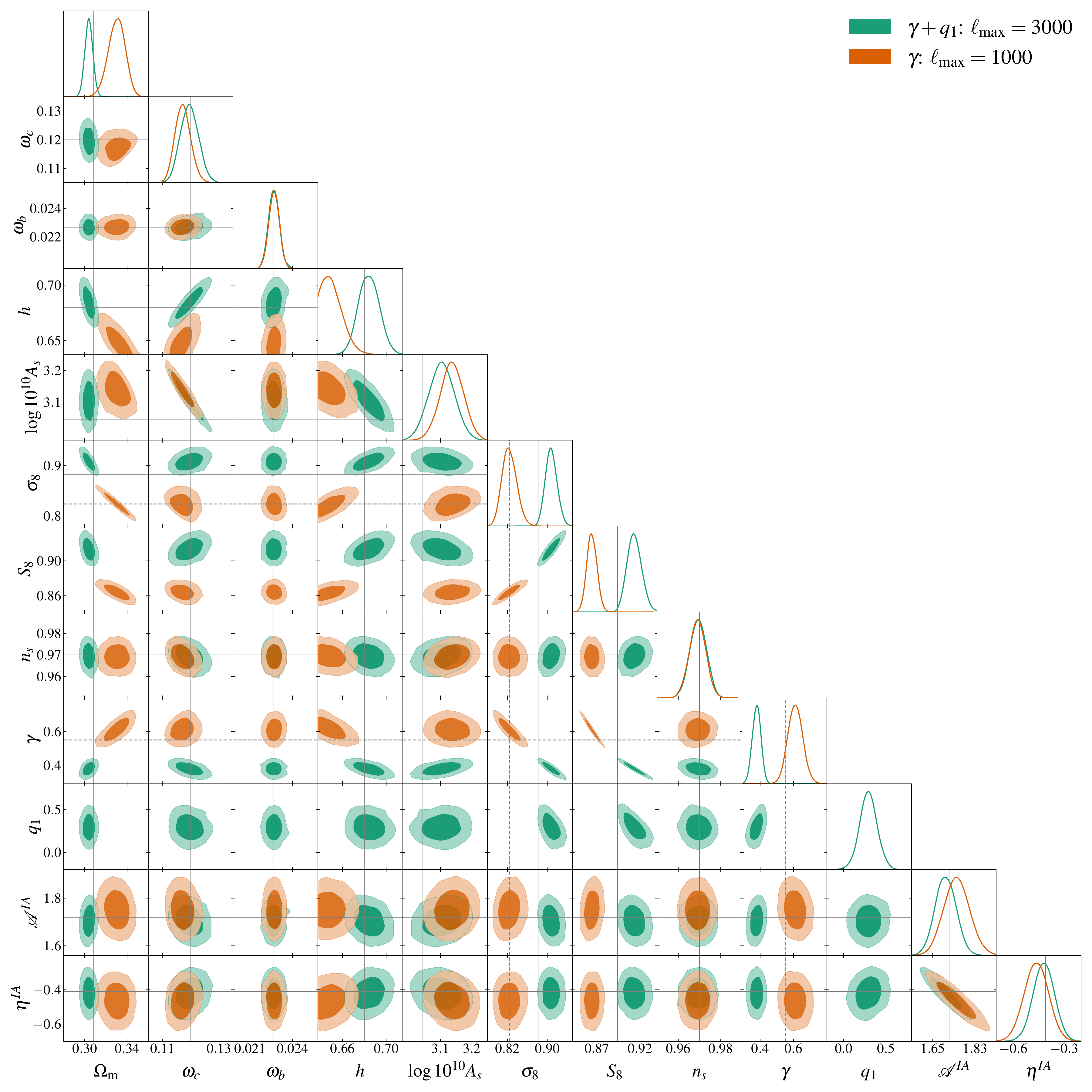}
    \caption{Full posterior distribution for showcasing the screening impact with nDGP mock data for a Stage-IV cosmic shear setup. Different colours correspond to the screened growth index model with $\ell_{\rm max}=3000$ (green) and the pseudo model with the modified linear growth and standard nonlinear structure formation with $\ell_{\rm max}=1000$ (orange). Solid grey lines mark the true values of the synthetic data, dashed grey lines mark the parameter values in the standard cosmology.}
    \label{fig:nDGP_Gmodels_full}
\end{figure}

In Fig.~\ref{fig:nDGP_Gmodels_full} we show the two-dimensional marginalised posteriors for the full
parameter space in $\gamma+q_1$ (green) and $\gamma$-pseudo (orange) models with with $\ell_{\rm max}=3000$ and $\ell_{\rm max}=1000$, respectively. The only significantly biased parameter is $\log{10^{10}~A_s}$, which is explained by a different time evolution of the linear growth factor in nDGP and the Linder gamma parametrisation. In Fig.~\ref{fig:MGrowths} we see that for values of $\gamma \sim 0.3-0.4$ we obtain an offset of  $D_\gamma/D_{\rm nDGP} \sim 0.97-0.98$. From the posterior maxima and the input fiducial values $A_s^{\rm mean}/A_s^{\rm fid}=1.06$ ($\Delta A_s = 6\%$). Therefore, lower (than in nDGP) values of the growth factor in the growth index model are compensated by a higher value of the primordial amplitude. We note that while the posterior-maxima of $\omega_c$ and $h$ are unbiased, their 2-dimensional contour demonstrates a $1\sigma$ bias that later propagates to a $1\sigma$ lower posterior maximum for $\Omega_{\rm m}$. We also found this bias when sampling in $\Omega_{\rm m}$ directly.

We notice that even for $\ell_{\rm max}=1000$, there is a significant bias in $h$ as well as in $\gamma$. Surprisingly, the pseudo $\gamma$ model finds a value of $\sigma_8$ at its GR value (a lower than expected value compared to nDGP), which is compensated by high values of $\Omega_{\rm m}$ ($1\sigma$ bias towards lower values in $\omega_c$ and $3\sigma$ bias towards lower values in $h$). 

\begin{figure}
    \centering
    \includegraphics[width=\columnwidth]{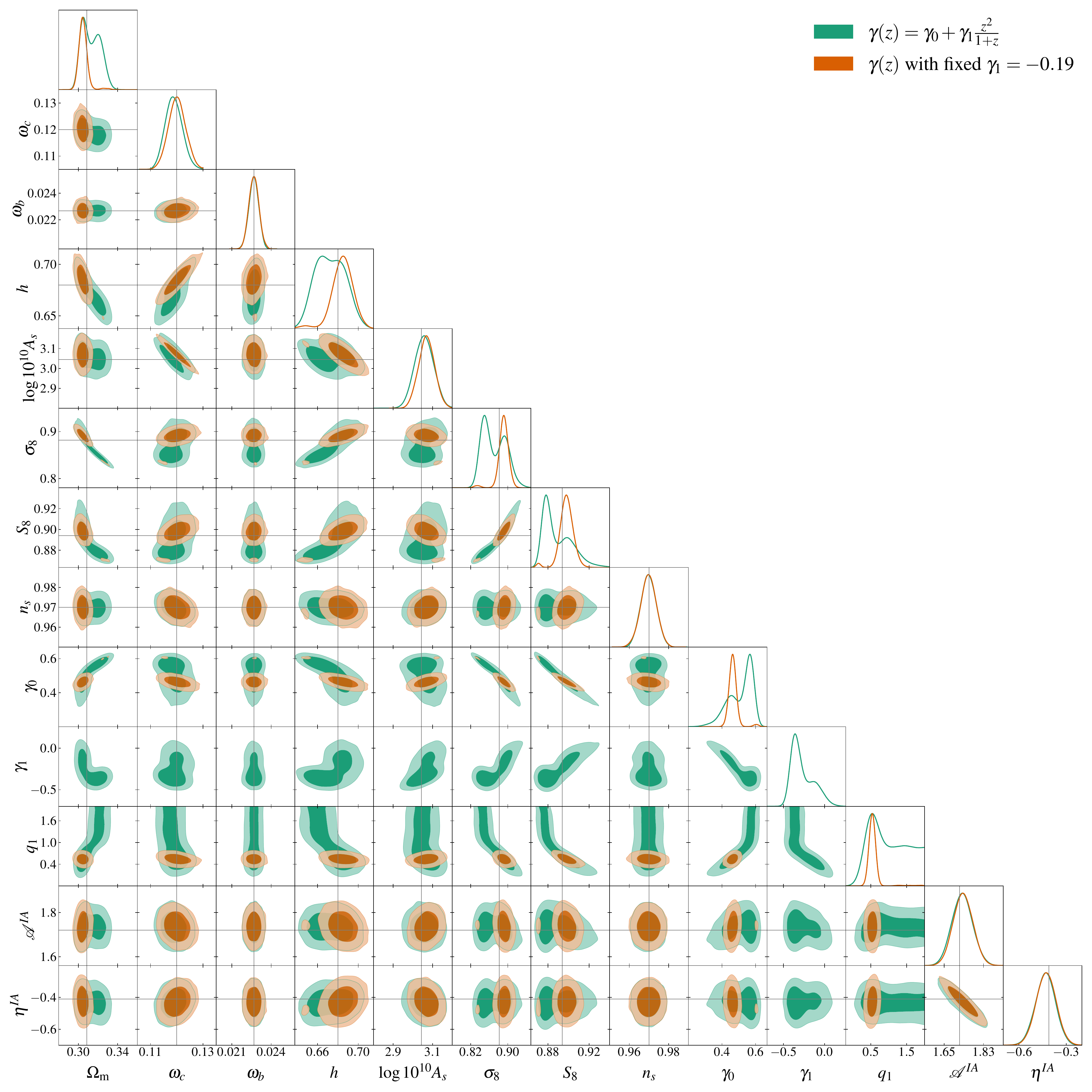}
    \caption{Full posterior distribution for tests with nDGP mock data for a Stage-IV cosmic shear setup with $\ell_{\rm max}=3000$. Different colours correspond to the time-dependent growth index model (green) and the same model but with the second coefficient fixed to $\gamma_1=-0.19$(orange). Grey lines mark the true values of the synthetic data.}
    \label{fig:nDGP_GQz_full}
\end{figure}

In Fig.~\ref{fig:nDGP_GQz_full} we show the two-dimensional marginalised posteriors for the full
parameter space in the $\gamma(z)+q_1$ model. Clearly, the bias towards higher values in $\log{10^{10}~A_s}$ vanishes due to a more accurate representation of the growth evolution. However, we notice significantly weaker constraints on $h$. This is expected from the degeneracy between the expansion rate and $\gamma$ (in this case $\gamma_0, \, \gamma_1$).  We show that this degeneracy is broken when $\gamma_1$ is fixed. 

\section{Other full posteriors}
\label{sec:FullPost}
\begin{figure}
    \centering
    \includegraphics[width=\columnwidth]{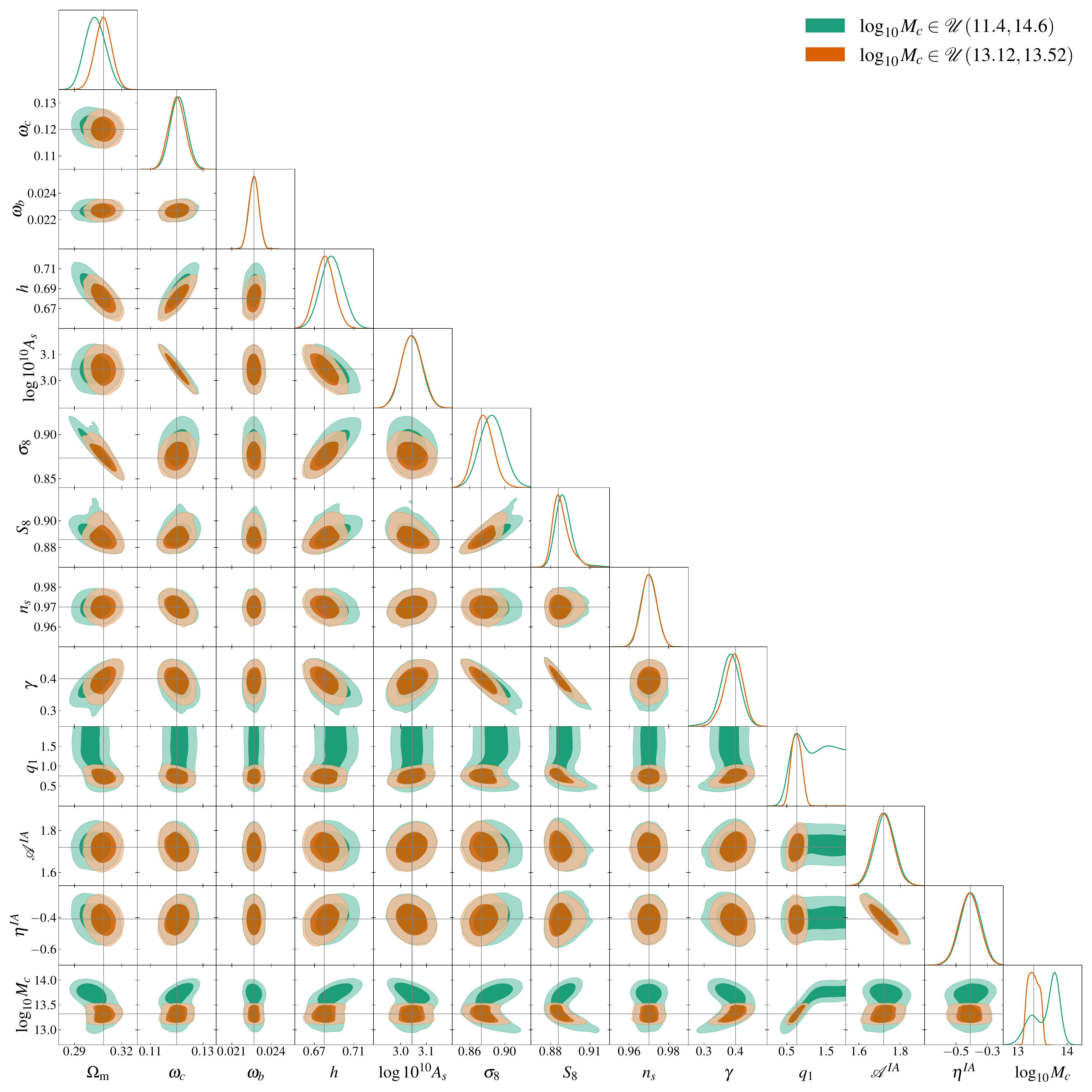}
    \caption{Full posterior distribution for investigating the degeneracy between the screening and baryonic parameters on $\gamma+q_1$ mock data for a Stage-IV cosmic shear setup with $\ell_{\rm max}=3000$. Different colours correspond to the flat broad priors on the baryonic feedback parameter $\log_{10}{M_c}$, (green) and a flat tight prior $\mathcal{U}(13.12, 13.52)$ on it (orange). Grey lines mark the true values of the synthetic data.}
    \label{fig:GQ_bar_full}
\end{figure}

\begin{figure}
    \centering
    \includegraphics[width=\columnwidth]{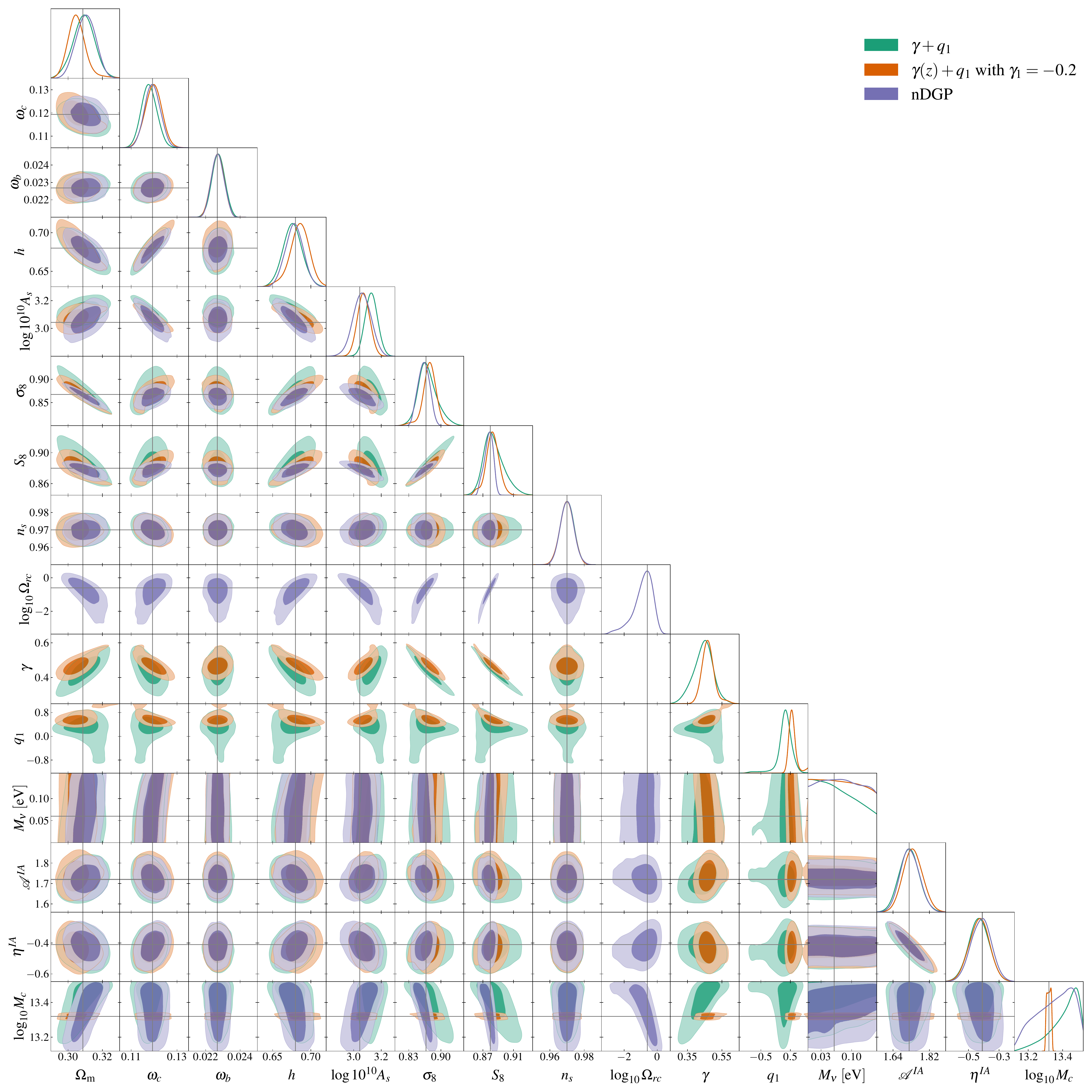}
    \caption{Full posterior distribution for the nDGP model with $\Omega_{rc}=0.25$ for a Stage-IV cosmic shear setup with $\ell_{\rm max}=3000$. Different colours correspond to the $\gamma+q_1$ model (green), the $\gamma(z)+q_1$ model with $\gamma_1=-0.2$ (orange), and the exact nDGP model (purple). Grey lines mark the true values of the synthetic data.}
    \label{fig:GQ_bar_Mnu_full}
\end{figure}

In Figs.~\ref{fig:GQ_bar_full} and \ref{fig:GQ_bar_Mnu_full} we show the two-dimensional marginalised posteriors for the full
parameter space in the analyses mentioned in the main text.


\end{document}